\documentstyle[eqsecnum,aps,graphics]{revtex}
\RequirePackage{ifthen}[1994/05/27]
\newboolean{drafton}	\setboolean{drafton}{false}
\ifthenelse{\boolean{drafton}}%
{    \newcommand{\equn}[1]{
        \mbox{}~\hspace*{\stretch{1}}~\begin{picture}(0,0)
           \setlength{\fboxsep}{.7mm}
	   \put(3,-12){\makebox[0mm][l]{\fbox{\small #1}}}
	\end{picture}
	\begin{equation}\label{#1}}
    \newcommand{\eqan}[1]{
	\setlength{\fboxsep}{.7mm}
        \mbox{}~\hspace*{\stretch{1}}~\begin{picture}(0,0)
	\put(3,-12){\makebox[0mm][l]{\fbox{\small #1}}}
	\end{picture}
	\begin{eqnarray}\label{#1}}}
{   \newcommand{\equn}[1]{\begin{equation}\label{#1}}
    \newcommand{\eqan}[1]{\begin{eqnarray}\label{#1}}}
\newcommand{\eqa}{\begin{eqnarray}}
\newcommand{\equ}{\begin{equation}}
\newcommand{\nuqe}{\end{equation}}
\newcommand{\uqe}{\end{equation}}
\newcommand{\naqe}{\end{eqnarray}}
\newcommand{\aqe}{\end{eqnarray}}
\newcommand{\nonu}{\nonumber}

\newcommand{\scs}{\scriptstyle}
\newcommand{\goto}{\rightarrow}
\newcommand{\half}{\frac{1}{2}}
\begin{document}
\twocolumn[\hsize\textwidth\columnwidth\hsize\csname@twocolumnfalse%
\endcsname
\draft

\title{Coupled Hamiltonians and \\Three Dimensional Short-Range Wetting Transitions}
\author{A.\ O.\ Parry and  P.\ S.\ Swain \cite{address}}
\address{Department of Mathematics, Imperial College \\ 180 Queen's Gate, London SW7 2BZ, United Kingdom.}
\maketitle

\begin{abstract}
We address three problems faced by effective interfacial Hamiltonian models of wetting based on a single collective coordinate $\ell({\bf y})$ representing the position of the unbinding fluid interface. Problems (P1) and (P2) refer to the predictions of non-universality at the upper critical dimension $d=3$ at critical {\it and} complete wetting respectively which are not borne out by Ising model simulation studies. (P3) relates to mean-field correlation function structure in the underlying continuum Landau model. Building on earlier work by Parry and Boulter we investigate the hypothesis that these concerns arise due to the coupling of order parameter fluctuations near the unbinding interface and wall. For quite general choices of collective coordinates $X_i({\bf y})$ we show that arbitrary two-field models $H \lbrack X_1,X_2 \rbrack$ can recover the required anomalous structure of mean-field correlation functions (P3). To go beyond mean-field theory we introduce a set ${\cal H}$ of Hamiltonians based on proper collective coordinates $s({\bf y})$ near the wall which have both interfacial and spin-like components. We argue that an optimum model $H \lbrack s,\ell \rbrack \in {\cal H}$, in which the degree of coupling is controlled by an angle  like variable $\delta^*$, best describes the non-universality of the Ising model and investigate its critical behaviour. For critical wetting the appropriate Ginzburg criterion shows that the true asymptotic critical regime for the local susceptibility $\chi_1$ is dramatically reduced consistent with observations of mean-field behaviour in simulations (P1). For complete wetting the model yields a precise expression for the temperature dependence of the renormalized critical amplitude $\theta$ in good agreement with simulations (P2). We highlight the importance of a new wetting parameter which describes the physics that emerges due to the coupling effects.
\end{abstract}

\vskip2pc]

\section{Introduction}
\subsection{Preliminary remarks}
During the last few decades the theory of continuous wetting transitions in inhomogeneous fluids and (simple) magnets has been extensively developed \cite{vol14,dietrich,edgar}. A striking feature of theory is the prediction that the critical exponents characterising the phase transition are sensitive to the details of the microscopic interactions (as well as fluctuation effects). Consider, for example, systems with long-range fluid-fluid and wall-fluid forces. Above the upper critical dimension, where mean-field (MF) theory is valid, the values of the critical exponents depend on the precise power law of molecular interactions \cite{dietrich}. Below the upper critical dimension, where fluctuations are important, one encounters weak and strong fluctuation regimes, with true universality only characteristic of the latter \cite{vol14,lf}. For systems with short-range forces the upper critical dimension for both critical and complete wetting transitions is $d=3$ \cite{lip}. Below this dimension we expect the critical exponents to take universal non-classical values at the respective transitions \cite{lf}. However at and above the upper critical dimension non-universality may appear in one of two ways. Firstly for $d>3$, corresponding to the MF regime, critical exponents may be non-universal if the decay lengths of the (exponential) wall-fluid and fluid-fluid forces are different \cite{ah}. More interesting is the behaviour at the upper critical dimension for which calculations based on the standard interfacial (or capillary-wave) model predict rather dramatic fluctuation-induced non-universality \cite{bhl,fh}. However, as is well known these predictions are not supported by extensive Monte Carlo (MC) simulations of critical wetting in the three dimensional Ising model which are instead consistent with MF theory \cite{blk,blw}. Therefore the status of the theory of wetting in systems with short-range forces at the upper critical dimension is somewhat controversial --- and has been so for over a decade.   

In recent years concerted effort has been made to refine effective Hamiltonian methods in an attempt to overcome this (and other) problem(s) of capillary-wave theory for three dimensional systems \cite{parryrev}. In addition developments in theory have been complemented by new Ising model simulation studies \cite{blf,blf1} which raise more issues and provide an independent test of the new ideas. In this paper we provide comprehensive details of a theory of criticality {\it and} correlation function structure at wetting transitions in systems with short-range forces based on a two-field effective Hamiltonian $H[s,\ell]$ which accounts for the coupling of order parameter fluctuations at the wall and depinning interface \cite{sp}. The model may be regarded as an extension of an earlier coupled theory due to Parry and Boulter (PB) \cite{pb,bp1,bp,pb1,bp2} which was largely restricted to the complete wetting phase transition. The new theory, described here also allows quantitative discussion of coupling effects at critical wetting as well as an improved treatment of the crossover from complete to critical wetting. Our analysis is consistent with the earlier PB theory which emerges naturally as a high temperature limit appropriate to modelling complete wetting. We shall argue that the coupled theory successfully explains three problems of (three dimensional) wetting theory including the critical wetting controversy mentioned above.     

As noted above our analysis is a further development of the PB theory which is itself a generalisation of a model and method introduced by Fisher and Jin (FJ) \cite{fjin,fjin2,jinf1,jinf2,fjp,pa}. While the FJ model does not allow for coupling effects, it does generalise the standard capillary-wave (CW) theory \cite{schick} by allowing for a position dependent stiffness coefficient which has important consequences.  We aim to keep our article as self-contained as possible and to continue our introduction we review these separate developments and focus on a number of important issues which provide more detail of the background and motivation for the present study.

\subsection{Microscopic models and the surface phase diagram}    
A suitable starting point for modelling adsorption in continuum fluids (or simple magnets) with short-range (contact) forces at a planar wall (situated in the $z=0$ plane) is the Landau-Ginzburg-Wilson (LGW) Hamiltonian \cite{nf,sulli}
\eqan{HLGW}
H_{LGW}[m] &=& \int d{\bf y} \Biggl\{ \int_{0}^{\infty} dz \left[ \frac{1}{2} (\nabla m)^2 + \phi(m) \right] \nonu \\
& & + \frac{c}{2} m_1^2 - m_1 h_1 \Biggr\}
\naqe
based on a local order parameter $m({\bf r}=({\bf y},z))$ which we will refer to as the magnetization. We will assume throughout this article that the continuum LGW model describes the same physics as the Ising model as regards wetting properties occurring above the roughening temperature $T_R$ of the latter (with $T_R \approx 0.54 T_C$ for a simple cubic lattice). The bulk potential function $\phi(m)$ yields coexistence between bulk phases $\alpha$ and $\beta$ at sub-critical temperatures $T<T_C$ and zero bulk field $h=0$. We shall refer to the bulk magnetizations for the up and down spin phases as $m_\alpha>0$ and $m_\beta<0$ respectively including cases in which the phase is metastable. The parameters $h_1$ and $c$ denote the surface field and enhancement respectively while 
\equ
m_1({\bf y}) \equiv m({\bf y},0) 
\uqe
is the magnetization at the wall. Often we shall abbreviate the surface term in (\ref{HLGW}) as $\phi_1(m_1)=\frac{c}{2} m_1^2 - h_1 m_1$. In most of our discussion of correlation functions we will let $k_BT=1$ for convenience but will explicitly include it in definitions of wetting parameters and for predictions of fluctuation effects.

\begin{figure}[h]
\begin{center}
\scalebox{0.7}{\includegraphics{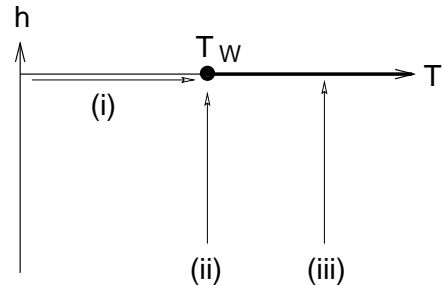}}
\end{center}
\caption{Section of the surface phase diagram. Along the cut in the $h=0$ axis the wall-$\beta$ interface is completely wet by the $\alpha$ phase. In Monte-Carlo simulation studies critical wetting has been investigated (for technical reasons) along path (ii), that is with $T=T_W$ and $h \rightarrow 0^-$.}
\label{pd}
\end{figure}

The model is expected to show wetting behaviour so that for sufficiently large surface field $h_1>h_1^W(>0)$ or temperature $T>T_W$ (with $T_W$ the wetting temperature) the wall-$\beta$ interface is completely wet by the $\alpha$ phase when $h=0^-$ \cite{sulli,pandit}. A section of a possible surface phase diagram is sketched in Fig.\ \ref{pd} and shows two types of continuous wetting transition. 
At each transition the mean thickness $\langle \ell \rangle$ of the adsorbed $\alpha$ phase diverges leading to large scale fluctuations in the position of the $\alpha \beta$ interface characterised by perpendicular and parallel correlation lengths $\xi_\perp$ and $\xi_\parallel$. We distinguish between critical wetting transitions which occur for $T \rightarrow T_W^-$ with fixed surface field (or for fixed temperature and $h_1 \rightarrow h_1^{W-}$) at $h=0^-$ (for example routes (i) and (ii) in Fig.\ \ref{pd}) and complete wetting in which $\langle \ell \rangle$ diverges for $T\ge T_W$ (or $h_1 \ge h_1^W$) as $h \rightarrow 0^-$ (shown as route (iii)). If we consider the surface tensions involved for the wall-$\alpha$ and wall-$\beta$ interfaces we can define a singular contribution $f_{\rm sing}$ for each transition by
\equn{fesum}
\sigma_{w\beta} = \sigma_{w\alpha} + \sigma_{\alpha \beta} + f_{\rm sing} 
\nuqe
where $\sigma_{\alpha \beta}$ is the free fluid interfacial tension between the bulk co-existing phases. The wall-$\beta$ interface is completely wet by the $\alpha$ phase if $f_{\rm sing}=0$ corresponding to zero contact angle \cite{dietrich}. 

\begin{figure}[h]
\begin{center}
\scalebox{0.55}{\includegraphics{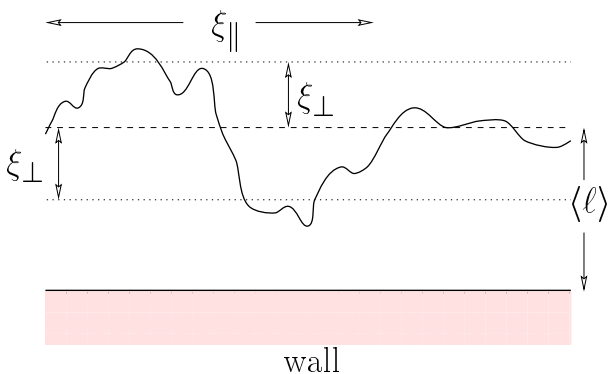}}
\end{center}
\caption{Schematic picture of the diverging lengthscales at a wetting transition.}
\label{fluc}
\end{figure}

For critical and complete wetting transitions occurring along routes (i) and (iii) respectively, the main critical exponents are defined by
\equ
\begin{array}{ccccc}
\langle \ell \rangle \sim \tau^{-\beta_s} & ; & \xi_\parallel \sim \tau^{-\nu_\parallel} & ; & f_{\rm sing} \sim \tau^{2-\alpha_s} 
\end{array} 
\uqe
and
\equ
\begin{array}{ccccc}
\langle \ell \rangle \sim |h|^{-\beta^{\rm co}} &;& \xi_\parallel \sim |h|^{-\nu_\parallel^{\rm co}} &;& f_{\rm sing} \sim |h|^{2-\alpha^{\rm co}} 
\end{array}
\uqe                                                     
where $\tau$ denotes the deviation from the critical wetting phase boundary
\equ
\tau \sim \left\{ \begin{array}{l} \frac{h_1^W-h_1}{h_1^W} \mbox{\hspace*{9mm} for fixed $T$} \\ \frac{T_W-T}{T_W} \mbox{\hspace*{10mm} for fixed $h_1$.} \end{array} \right.
\uqe
A schematic illustration of the various lengthscales in the problem is shown in Fig.\ \ref{fluc}. 

In principle the transverse and perpendicular correlation lengths can be calculated from the pair correlation function
\equn{LGWcf}
G({\bf r}_1,{\bf r}_2) = \langle m({\bf r}_1) m({\bf r}_2) \rangle - \langle m({\bf r}_1) \rangle \langle  m({\bf r}_2) \rangle 
\nuqe
or its transverse Fourier transform (with ${\bf y}_{12} = {\bf y}_2-{\bf y}_1$)
\equ
G(z_1,z_2;{\bf q}) \equiv \int d {\bf y}_{12} {\rm e}^{i {\bf q}.{\bf y}_{12}} G({\bf r}_1,{\bf r}_2)
\uqe
for particle positions near the $\alpha \beta$ interface. 

In effective Hamiltonian theories based on a collective coordinate $\ell({\bf y})$ representing the local height of the interface (at vector position ${\bf y}$ along the wall) $\xi_\parallel$ may be calculated from the analogous height-height correlation function while the root mean square fluctuation is $\xi_\perp=\Bigl[ \langle \ell^2 \rangle - \langle \ell \rangle^2 \Bigr]^{1/2}$. In fact the two lengths are not independent of each other and in $d=3$ effective Hamiltonian theories predict (see \cite{vol14,dietrich,lf} and also later)
\equn{gas}
(\kappa \xi_\perp)^2 = \omega \ln \Bigl( 1+(\Lambda \xi_\parallel)^2 \Bigr) 
\nuqe
with $\kappa$ the inverse (true) correlation length of the bulk $\alpha$ phase and $\Lambda$ the high momentum cut-off. Here $\omega$ is the wetting parameter
\equ
\omega \equiv \frac{k_B T \kappa^2}{4 \pi \Sigma_{\alpha \beta}}
\uqe
with $\Sigma_{\alpha \beta}$ the stiffness (related to the tension) of the free $\alpha \beta$ interface. Importantly $d=3$ is the upper critical dimension for interfacial roughness (for thermal fluctuations) and for $d>3$ the interface width $\sim \xi_\perp$ is only of order a bulk correlation length.
     
A direct analysis of wetting in the LGW model is only possible in MF approximation  \cite{nf,pandit} which ignores fluctuation effects and is equivalent to an effective value of the wetting parameter equal to zero. Minimising (\ref{HLGW}) determines the MF free energy
\equ
{\cal F}_{MF} = \min H_{LGW}[m]
\uqe
and yields a standard Euler-Lagrange equation for the MF profile $\tilde{m}(z)$. For our purposes we note only that for $c>\kappa$ \cite{pandit} sections of the MF surface phase diagram are similar to Fig.\ \ref{pd} with the wetting field and temperature determined by the relation
\equ
\begin{array}{lcl}
h_1^W = c m_\alpha(T^{MF}_W) &;& h=0^-
\end{array}
\uqe

The MF values of the critical singularities at critical and complete wetting are                                                       
\equn{mf1}
\begin{array}{ccccc}
\kappa \langle \ell \rangle \sim \ln \tau^{-1} &;& \xi_\parallel \sim \tau^{-1} &;& f_{\rm sing} \sim \tau^2 
\end{array}
\nuqe
and
\equn{mf2}
\begin{array}{ccccc}
\kappa \langle \ell \rangle \sim \ln |h|^{-1} &;& \xi_\parallel \sim |h|^{-1/2} &;& f_{\rm sing} \sim h \ln |h| 
\end{array}
\nuqe                                                       
respectively. Substitution of the MF critical exponents into the hyperscaling relations $2-\alpha_s=(d-1) \nu_\parallel$ and $2-\alpha_s^{\rm co}=(d-1) \nu_\parallel^{\rm co}$ determines the upper critical dimensions as $d^*=d^*_{\rm co}=3$ as quoted earlier \cite{vol14,dietrich,lip}. It is therefore possible that fluctuations alter the value of critical exponents and amplitudes in $d \le 3$ but to understand this we need to resort to effective Hamiltonian methods.

\subsection{The capillary-wave model}    
The fluctuation theory of wetting is based almost entirely on the phenomenological CW model \cite{vol14}
\equn{inter}
H[\ell] = \int d{\bf y}\left\{ \frac{1}{2} \Sigma_{\alpha \beta} (\nabla \ell)^2 + W(\ell) \right\}
\nuqe
This has been generally accepted as a suitable coarse-grained description of an interface whose fluctuations occur on lengthscales much larger than the bulk correlation length (hence the cut-off restriction $\Lambda \ll \kappa$ in (\ref{gas})). Instead of a microscopic order parameter the Hamiltonian is a functional of the collective coordinate $\ell({\bf y})$ mentioned earlier. Long wavelength fluctuations in the local position of the interface increase its surface area and are resisted by the stiffness term in (\ref{inter}). The direct or bare interaction of the interface with the wall is mediated by the binding potential $W(\ell)$ constructed with reference to MF approaches (see later). For systems with short-range forces such as the LGW model the form usually assumed for modelling continuous wetting transitions is \cite{bhl,schick}
\equn{W}
W(\ell) = \bar{h} \ell + 2 \kappa m_\alpha \tau {\rm e}^{-\kappa \ell} + b {\rm e}^{-2 \kappa \ell} 
\nuqe
together with a hard wall contribution which restricts configurations to $\ell({\bf y})>0$. Here $\bar{h}=(m_\beta-m_\alpha) h$ is a measure of the deviation from two-phase coexistence while $b$ is a positive constant near $T_W$ (which is required to ensure the stability of the transition). The deviation from the critical wetting phase boundary, $\tau$, is given explicitly as
\equ
\tau=\frac{h_1-cm_\alpha}{c+\kappa}
\uqe
and implies that the overall sign of the leading exponential is positive for $T>T_W^{MF}$. Note that by minimising the binding potential we recover the MF expressions for the mean film thickness and free energy (\ref{mf1}, \ref{mf2}) quoted earlier. Also note that for complete wetting only the first exponential term is required to model the transition. 

If we assume that the fluctuations $\delta \ell({\bf y}) = \ell({\bf y}) - \tilde{\ell}$ (where $\langle \ell \rangle = \tilde{\ell}$ in MF theory) are small (which is valid for $d>3$) then it is straightforward to calculate the structure factor
\equ
S({\bf q}) = \int d{\bf y}_{12} {\rm e}^{i {\bf q}.{\bf y}_{12}} \langle \delta \ell({\bf y}_1) \delta \ell({\bf y}_2) \rangle
\uqe
which has a simple Lorentzian form
\equ
S({\bf q}) = \frac{k_B T}{W'' + q^2 \Sigma_{\alpha \beta}}
\uqe
reminiscent of Ornstein-Zernike theory. Here $W'' \equiv \left. \frac{d^2 W}{d \ell^2} \right|_{\tilde{\ell}}$ which allows us to identify the transverse correlation $\xi_\parallel^2 = \frac{\Sigma_{\alpha \beta}}{W''}$ \cite{lip2}. If we assume that small fluctuations $\delta \ell$ correspond to magnetization fluctuations which translate the MF profile $\tilde{m}(z)$ we are led to the prediction \cite{henderson}
\equn{mfident}
G(z_1,z_2;{\bf q}) \approx \tilde{m}'(z_1) \tilde{m}'(z_2) S({\bf q}) 
\nuqe
for the MF correlation function for positions $z_1, z_2$ close to the $\alpha \beta$ interface. Importantly this agrees with the explicit solution of the Ornstein-Zernike equation for the LGW model (in MF approximation) which reads \cite{evans} (setting $k_BT=1$)
\equn{ozdif}
\Bigl( -\frac{\partial^2}{\partial z_1^2} + \phi''(\tilde{m}(z_1)) + q^2 \Bigr) G(z_1,z_2;{\bf q}) = \delta(z_1-z_2)
\nuqe
As we shall see in the next sections MF identifications such as (\ref{mfident}) can be made precise for arbitrary $z_1, z_2$ using generalized effective Hamiltonian and collective coordinate theories. 

For $d<3$, renormalization group (RG) and transfer matrix studies of the CW model with binding potential (\ref{W}) predict universal critical behaviour for both transitions \cite{vol14,lf,burkhardt}. For these dimensions the precise form of $W(\ell)$ is not essential because the critical properties are determined by fixed points of the RG transformation. Thus in $d=2$ approximating $W$ by a square well  recovers the same critical exponents for critical wetting as those found in the exact Ising model calculation \cite{ab}. However for $d \ge 3$ the precise structure of  $W(\ell)$ is essential because there is no non-trivial fixed point Hamiltonian \cite{vol14}.     

At the upper critical dimension $d=3$, linear RG calculations predict that the critical singularities are non-universal depending on the wetting parameter $\omega$ \cite{bhl,fh}. For critical wetting the results are particularly dramatic, for example the correlation length exponent is given by
\equn{nu}
\nu_\parallel = \left\{ \begin{array}{lcl} (1-\omega)^{-1} & & \mbox{\hspace*{3mm} for $0 \le \omega < \frac{1}{2}$} \\ (\sqrt{2} -\sqrt{w})^{-2} & & \mbox{\hspace*{3mm} for $\frac{1}{2} < \omega <2$} \\ \infty & & \mbox{\hspace*{3mm} for $\omega > 2$} \end{array} \right.
\nuqe
where the last regime corresponds to an exponentially fast divergence. Note also that for $\omega<2$ the wetting temperature is unaltered so that $\tau=0$ still denotes the critical wetting phase boundary. For complete wetting the predictions are less dramatic and critical exponents retain their MF values \cite{fh}. Nevertheless critical amplitudes are renormalized and of particular interest is the dimensionless adsorption amplitude
\equn{theta}
\theta \equiv \stackrel{\rm lim}{\scs \bar{h} \rightarrow 0} \left\{ \frac{\kappa \langle \ell \rangle}{\ln |\bar{h}|^{-1}} \right\}
\nuqe
which is given by \cite{fh}
\equn{thetacw}
\theta = \left\{ \begin{array}{lcl} 1+\frac{\omega}{2} & & \mbox{\hspace*{3mm} for $0 < \omega < 2$} \\ \sqrt{2\omega} & & \mbox{\hspace*{3mm} for $\omega >2$} \end{array} \right.
\nuqe

While these predictions are made using an approximate linear RG scheme which cannot handle the hard-wall contribution to $W(\ell)$ precisely, there are good reasons for believing that the results are exact at least in the important regime $\omega<2$. In particular they are supported by MC simulations of a discretized CW model \cite{gk} and also by numerical studies of a non-linear RG analysis which does allow for the hard-wall term \cite{lf}.
     
The value of the wetting parameter is not determined by the CW theory and must be regarded as input into the model. During the time since these predictions were first made the temperature dependence $\omega(T)$ for the simple cubic Ising model has been studied in some detail \cite{fishw}. Near $T_R$, $\omega \approx 0.5$ but its value rises sharply on increasing $T$ and is very close to $\omega \approx 0.8$ for all temperatures in the range $0.6 T_C$--$T_C$. On approaching the bulk critical point, hyperuniversality implies that the wetting parameter tends to a universal value
\equ
\stackrel{\rm lim}{\scs T \rightarrow T_C^{-}} \Bigl\{ \omega(T) \Bigr\} \equiv \omega_C
\uqe
estimated as $\omega_C \approx 0.77_5$. Thus for the Ising model RG calculations based on the CW model predict the numerical values
\equ
\left. \begin{array}{lcl}
\nu_\parallel &\approx& 3.7 \\
\theta &\approx& 1.4 \end{array} \right\} \mbox{\hspace*{1mm} CW theory}
\uqe        
(assuming $0.6 T_C \lesssim T < T_C$) significantly different from the MF results (\ref{mf1}, \ref{mf2}).

\subsection{The Fisher-Jin model and position dependent stiffness}   
In a series of important papers \cite{fjin,fjin2,jinf1,jinf2,fjp} FJ reassessed the status of the CW model and suggested specific modifications that have some interesting consequences in $d=3$. They were principally motivated by one of the central problems of wetting theory (see (P1) below) and forwarded a novel explanation. In retrospect however there are further problems that need to be addressed that the FJ model cannot provide answers for and it is our view that the FJ analysis should be regarded as an important step towards a fuller description of fluctuation effects rather than an end in itself. Nevertheless the systematic approach which they adopted for integrating out degrees of freedom is crucially important to subsequent theoretical developments and we recall some of the essential features of their formalism.

FJ seek to derive an interfacial Hamiltonian from a LGW model and begin their analysis by emphasizing the need to carefully define the collective coordinate $\ell({\bf y})$. There is of course a good deal of freedom in the choice of definition \cite{fjin} (see later) and utility is an important criterion. By far the easiest option is to adopt the crossing criterion in which $\ell({\bf y})$ corresponds to the surface of fixed magnetization $m^X$. In their analysis FJ set $m^X=0$ but in keeping with subsequent developments we keep $m^X$ arbitrary. The generalized effective Hamiltonian for the surface of fixed magnetization is defined as \cite{fjin}
\equ
{\rm e}^{-H_{FJ}[\ell;m^X]} = \int_C {\cal D}m \; {\rm e}^{-H_{LGW}[m]}
\uqe
where $C$ denotes a constrained functional integral or partial trace over configurations $m({\bf r})$ which respect the crossing criterion
\equn{CC}
m \Bigl( {\bf r} = ({\bf y},\ell({\bf y})) \Bigr) = m^X
\nuqe

Next FJ argue that for a given collective coordinate configuration all other fluctuations are small and may be ignored. This leads to a saddle point identification 
\equ
H_{FJ}[\ell;m^X] = H_{LGW}[m_\Xi({\bf r};\ell)]
\uqe
where $m_\Xi$ is the magnetization configuration which minimises $H_{LGW}[m]$ subject to the crossing criterion. Fortunately in order to derive effective Hamiltonians describing long wavelength fluctuations it is sufficient to consider planar constrained profiles $m_\pi(z;\ell_\pi)$ which satisfy the Euler-Lagrange equation \cite{fjp,nap}
\equn{el}
m_\pi''(z) = \phi' \Bigl( m_\pi(z) \Bigr)
\nuqe
with boundary conditions
\eqa
\left. \frac{\partial m_\pi}{\partial z} \right|_{z=0} &=& \phi_1'(m_{\pi1}) \\
\stackrel{\lim}{\scs z \rightarrow \infty} m_\pi(z) &=& m_\beta
\aqe
(where $m_{\pi1}=m_\pi(0;\ell_\pi)$) and the planar constraint
\equ
m_\pi(\ell_\pi;\ell_\pi) = m^X
\uqe
                                                   
In this way FJ derive
\eqan{HFJ}
H[\ell;m^X] &=& \int d{\bf y} \Biggl\{ \frac{1}{2} \Sigma \Bigl( \ell({\bf y});m^X \Bigr) \Bigl[ \nabla \ell({\bf y}) \Bigr]^2 \nonu \\
& & + W \Bigl( \ell({\bf y});m^X \Bigr) \Biggr\} 
\naqe
where the binding potential and position dependent stiffness for the surface of fixed magnetization $m^X$ are determined by $m_\pi(z;\ell_\pi)$ as
\eqa
\lefteqn {W \Bigl( \ell({\bf y});m^X \Bigr) = \Biggl\{ \phi_1(m_{\pi1})} \nonu \\
 & & \left. + \int_0^{\infty} dz \left[\frac{1}{2} \left( \frac{\partial m_\pi}{\partial z} \right)^2 +  \phi(m_\pi) \right] \Biggr\} \right|_{\ell_\pi=\ell({\bf y})}
\aqe
(ignoring some $\ell$ independent terms) and
\eqa
 \Sigma \Bigl( \ell({\bf y});m^X \Bigr) &=&  \left. \int_0^\infty dz  \left( \frac{\partial m_\pi}{\partial \ell_\pi} \right)^2  \right|_{\ell_\pi=\ell({\bf y})} \nonu \\
&=& \Sigma_{\alpha \beta} + \Delta \Sigma \Bigl( \ell({\bf y});m^X \Bigr) 
\aqe                                                  
Note that the position dependent increment $\Delta \Sigma \goto 0$ as $\ell \rightarrow \infty$ for sensible choices $m_\alpha >m^X>m_\beta$ corresponding to surfaces of fixed magnetization which unbind from the wall. For these choices the explicit expansion of the binding potential is very similar to the CW expression (\ref{W}) but the $\Delta \Sigma$ term is new \cite{fjin2}
\equn{fjstiff}
\Delta \Sigma(\ell;0) = 2 \kappa m_\alpha \tau {\rm e}^{-\kappa \ell} - 2\kappa^2 m_\alpha^2 \nu \kappa \ell {\rm e}^{-2\kappa \ell} + \cdots
\nuqe
with $\nu=\frac{c-\kappa}{c+\kappa}$. Note that by construction, minimization of $W(\ell;m^X)$ recovers the MF position of the surface of fixed magnetization $m^X$.

The (linear) RG analysis of the FJ model is more complicated than the CW model because the flows of $\Delta \Sigma$ and $W$ are coupled. Nevertheless FJ show \cite{fjin2,jinf1} that the effective potential (with $t \equiv \ln b$ the usual infinitesimal rescaling factor)
\equ
W_{\rm eff}^{(t)}(\ell) \equiv W^{(t)}(\ell;0) + \frac{\omega \Lambda^2}{2\kappa^2} (1-{\rm e}^{-2t}) \Delta \Sigma^{(t)}(\ell;0) 
\uqe
satisfies the same diffusion-type flow equation familiar from the linear RG analysis Fisher and Huse made of the CW model \cite{fh}. Due to the negative next-to-leading order exponential term in $\Delta \Sigma(\ell;0)$ the critical wetting transition is destabilised for sufficiently small $\omega<\omega^*$ and is replaced by a weakly first-order phase transition. FJ argue that the value of $\omega^*$ is in the range $\frac{1}{2} < \omega^* <1$ for the Ising model although recent non-linear RG studies of the FJ model estimate a significantly larger value $\omega^* \approx 2$ \cite{B}. For $\omega>\omega^*$ the critical wetting behaviour is similar to the CW model.      

The FJ description of the complete wetting transition is the same (at leading order) as the CW model because only the first exponential term in $W_{\rm eff}^{(t)}(\ell)$ is required. Thus their model predicts the same value of the adsorption critical amplitude $\theta$ as CW theory, that is (\ref{thetacw}).

\subsection{Three problems of capillary wave theory}      
Here we point out three problems faced by the standard CW theory of wetting. Two relate directly to discrepancies with Ising model simulation studies and are associated with fluctuation effects at the upper critical dimension $d=3$. The third concerns questions of self consistency and arises when we compare the CW theory with MF studies of correlation function structure in the LGW model. We will argue, building on the earlier analysis of PB that the final problem provides the key to understanding the other two.

\newcounter{c}
\begin{list}{\bf (P\arabic{c})}{\usecounter{c} }
\item {\bf Ising simulations of critical wetting} \\     
Extensive MC simulations of wetting transitions in the $d=3$ Ising model appear to show that the critical wetting transition is characterised by MF-like critical exponents. Specifically Binder, Landau and co-workers \cite{blk,blw} studied two wetting transitions occurring near $0.6T_C$ and $0.9T_C$ (corresponding to different choices of the surface field) and establish the divergence of the surface susceptibility $\chi_1 \equiv \frac{\partial m_1}{\partial h}$. According to scaling expectations $\chi_1$ should behave as \cite{blk}
\equ
\chi_1 = \tau^{-1} \chi(h\tau^{-2\nu_\parallel})
\uqe
leading to the prediction $\chi_1 \sim |h|^{-\frac{1}{2\nu_\parallel}}$ along the critical isotherm (route (ii) in Fig.\ \ref{pd}). Contrary to the CW model the Ising data are rather well fitted by the MF result $\chi_1 \sim |h|^{-1/2}$ corresponding to $\nu_\parallel=1$.
        
Despite widespread discussion in the literature only two detailed explanations have been proposed. Shortly after the simulation results were reported Halpin-Healey and Br\'{e}zin \cite{hb} used the CW model to calculate Ginzburg criteria for various response functions. For the surface susceptibility $\chi_1$ they show that crossover from MF to non-classical behaviour begins when the transverse correlation length is close to a value satisfying
\equn{bh}
\omega \left[ \frac{1}{2} \ln \left( 1+ \xi_\parallel^2 \Lambda^2 \right) + \frac{1}{1+\xi_\parallel^2 \Lambda^2} -1 \right] = 1
\nuqe
Typically this yields a value for $\xi_\parallel$ of a few bulk correlation lengths which Halpin-Healey and Br\'{e}zin argued was not quite attained in the MC work. However subsequent simulations for larger lattices and smaller bulk fields still show no substantial deviation from MF behaviour \cite{blw}. Even re-analysis of the data for the susceptibility critical amplitude (which is extremely sensitive to fluctuation effects) shows only minor deviations from classical expectations and is wildly different from CW theory \cite{peb}. Moreover simulations of a discretized CW model shows no discernible MF regime and are supportive of the RG predictions \cite{gk,gkl}. 

The second suggestion is the stiffness instability mechanism of FJ who argue that the Ising model wetting transition is weakly first-order \cite{fjin2}. While this may very well be the case (and certainly remains a possibility in our theory) on its own it does not provide a quantitative explanation of why the MC data for $\chi_1$ is MF-like.

\item {\bf Ising model simulations of complete wetting} \\  
Recently Binder, Landau and Ferrenberg (BLF) \cite{blf,blf1} have studied thin-film Ising models with opposite surface fields $h_1$, $h_D=-h_1$ acting on the spins in the planes $z=0, D$ respectively. The phase diagram for this system was originally predicted by Parry and Evans \cite{pe} on the basis of a MF analysis of the appropriate LGW model and shows a number of intriguing features. Of particular interest is the behaviour of the correlation length in the temperature window $T_W<T<T_C$ (and $h=0$) which probes finite-size effects at the complete wetting transition. In the thin-film this corresponds to the one-phase regime in which an $\alpha \beta$ interface sits at the centre of the slab and is subject to very large scale fluctuations. The transverse correlation length is predicted to be exponentially large in the width $D$ reflecting the shallowness of the effective binding potential. It is a simple matter to construct a CW theory for this system and use a linear RG analysis to predict modifications to the MF result. This yields \cite{bp1,bp}
\equn{oppos}
\xi_\parallel \sim {\rm e}^\frac{\kappa D}{4\theta} \mbox{\hspace*{1cm} for $\kappa D \rightarrow \infty$}
\nuqe
where $\theta$ is the complete wetting adsorption critical amplitude introduced earlier. The prediction of an exponentially large correlation length was checked and confirmed by BLF for several temperatures in the range $T_W<T<T_C$ (it can be extracted relatively easily from the the total and mid-point susceptibility). However the exponent of the exponential term is inconsistent with both MF and CW theory! Using their data Parry, Boulter and Swain (PBS) \cite{pbs} have extracted the effective value of the critical amplitude $\theta$ as a function of temperature. A plot of the simulation results is shown in Fig.\ \ref{bigtheta} and shows a substantial increment to the expected value $\theta_{CW} \approx 1.4$ (recall $\theta_{MF} \approx 1$) for temperatures deep in the complete wetting regime ($T>T_W\approx 0.9T_C$). 
\begin{figure}[h]
\begin{center}
\scalebox{0.42}{\includegraphics{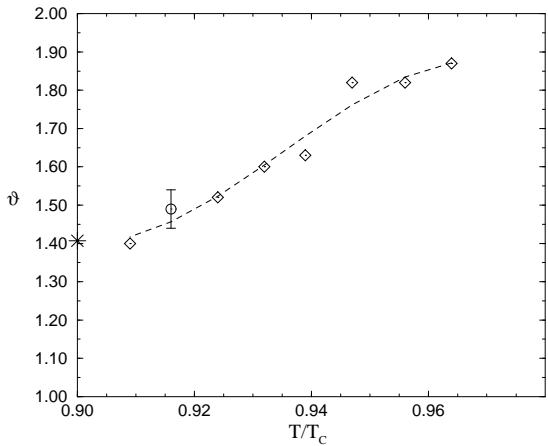}}
\end{center}
\caption{BLF data for $\theta$ vs.\ $T/T_C$ showing the anomalous enhancement to the CW result $\theta=1+\omega/2 \approx 1.4$ deep in the complete wetting regime. Apart from the point at $T/T_C = 0.916$ the error is within the symbol. The dotted line is a cubic fit to the data while the asterix is the extrapolated value of $\theta$ at $T_W \approx 0.9 T_C$. The decrease in the value of $\theta$ as $T \goto T_W^+$ is indicative of novel crossover behaviour near critical wetting associated with coupling effects.}
\label{bigtheta}
\end{figure}
Specifically at high temperatures the simulation results are consistent with $\theta \approx 1.7$--$1.8$. The abnormally large values of $\theta$ are inexplicable using the CW and FJ models and raise further doubts whether the standard picture of interfacial fluctuation effects captures all the essential physics at the upper critical dimension.

\item {\bf Correlation function structure}\\  
While the importance of the two problems described above can be recognised immediately it is not at all clear how to modify existing effective Hamiltonian theory to overcome them (assuming that this is where the fault lies). However there is a third long-standing puzzle associated with wetting theory, which while less dramatic than the questions concerning non-universality, focuses attention on a specific defect of any model based on a single collective coordinate $\ell({\bf y})$.    

MF studies of the order parameter correlation function $G({\bf r}_1,{\bf r}_2)$ for a number of different density functional models of wetting reveal intriguing features for positions close to the wall which cannot be described using the CW model (or for that matter the amended model proposed by FJ) \cite{pb2,pe2}. It should be emphasised that the MF studies are consistent with some curious exact sum rule requirements which link correlation function and thermodynamic singularities and which also defy explanation using standard arguments \cite{pe2,hvs}. We expect therefore that much of the correlation function structure described below is also present in the full three dimensional LGW model beyond MF approximation.   

First, consider the complete wetting transition in the LGW model. The MF correlation function satisfies the differential equation (\ref{ozdif}) whose solution has a simple Lorentzian form for $z_1,z_2$ near the $\alpha \beta$ interface. However for particle positions at (or close to) the wall the solution is non-Lorentzian \cite{pb2}
\eqan{PBresult}
\lefteqn{G(0,0;q) \approx} \nonu \\ 
& & \frac{\tilde{m}_1^{'2}} {(c+\kappa)\tilde{m}_1^{'2} + q^2 \left[ \sigma_{w\alpha} - \phi_1(\tilde{m}_1) + \frac{\sigma_{\alpha \beta} + f_{\rm sing}}{1+q^2 \xi_\parallel^2} \right]}
\naqe
where $\tilde{m}'_1$ is the gradient of the magnetization at the wall (which is of order $\tau$ and may be regarded a constant). As emphasised by PB the wave-vector dependence shows crossover from coherent to intrinsic behaviour depending on the scaling variable $x \equiv q \xi_\parallel$. By this we mean that for finite $\xi_\parallel$ and $q \rightarrow 0$ the effective coefficient of  $q^2$ in the denominator is the full surface tension (or more properly, stiffness) of the wall-$\beta$ interface and indicates the coherence of asymptotically long wavelength fluctuations in the wetting film. This manifests itself in the second moment, defined in the expansion $G(z_1,z_2;{\bf q}) = \sum_{n=0}^\infty q^{2n} G_{2n}(z_1,z_2)$, such that
\equn{sumr}
-G_2(0,0) \propto \sigma_{w\beta} - \phi_1(\tilde{m}_1)
\nuqe
This is no artifact of the MF approximation since an equivalent statement is known to be precisely true for the phenomenon of complete drying by a vapour phase ($\beta$) at a hard wall-liquid ($\alpha$) interface. The exact sum-rule reads \cite{hvs}
\eqan{g2decay}
-G_2(0,0) &=& \sigma_{w\alpha} + \sigma_{\alpha \beta} + f_{\rm sing} \nonu \\
& = & \sigma_{w\beta}
\naqe
and is valid for arbitrary fluid-fluid forces. 

On the other hand in the scaling limit of finite $q$ and $\xi_\parallel \rightarrow \infty$ (i.e.\ $h\rightarrow 0^-$) the correlation function reduces to the appropriate expression for fluctuations typical of the wall-$\alpha$ interface. For this case the effective coefficient of $q^2$ in the denominator is the local, intrinsic stiffness $\Sigma_{w\alpha}=\sigma_{w\alpha}-\phi_1(\tilde{m}_1)$.        

The correlation function with one particle at the wall and one at the $\alpha \beta$ interface also shows interesting features. For example in their MF study of drying at a hard-wall within the Sullivan density functional model Parry and Evans derive \cite{pe2}
\equ
-\frac{G_2(0,\tilde{\ell})}{\tilde{m}'(\tilde{\ell})} \approx \xi_\parallel^2 + \frac{\tilde{\ell}}{\kappa} + \frac{\sigma_{w\beta}}{\tilde{m}'_1}
\uqe
where here $\tilde{m}'(\tilde{\ell})$ denotes the gradient of the equilibrium number density at the vapour-liquid ($\alpha \beta$) interface $z = \tilde{\ell}$. The second and third terms on the r.h.s.\ correspond to logarithmic next-to-leading order and coherent properties, respectively. The same behaviour emerges in the LGW model at MF level although, unfortunately, there is no exact sum-rule to compare with.       

A cause for concern for theorists is that hardly any of these features are describable using one-field Hamiltonians such as the CW and FJ models. In fact one of the virtues of the FJ approach is that one may unambiguously calculate a correlation function $G^{FJ}(z_1,z_2;{\bf q})$ (for a given $m^X$) at MF level (and beyond) since the theory allows a one-to-one connection between order parameter and collective coordinate configurations through the field $\ell({\bf y})$. In this way it is straightforward to show that the FJ model with $m^X=0$ (and by implication the CW model) cannot recover the non-Lorentzian, next-to-leading order and coherent behaviour manifest in the MF $G(z_1,z_2;{\bf q})$ at complete wetting \cite{pa}. 

Similar remarks apply to the critical wetting transition. Most importantly the correlation function is non-Lorentzian at the wall --- which can be read from (\ref{PBresult}) on noting that $\tilde{m}'_1 \sim \tau$ and $\sigma_{w\alpha} - \phi_1(\tilde{m}_1) \sim \tau^2$. The result is conveniently written 
\equn{Gcw} 
\frac{1}{G^{w\beta}(0,0;{\bf q})} \approx \frac{1}{G^{w\alpha}(0,0;{\bf q})}  + \frac{\rm const.}{1+q^2 \xi_\parallel^2}
\nuqe 
and shows the necessary crossover from singular to intrinsic behaviour dependent on the scaling variable $x=q\xi_\parallel$. This expression is consistent with a number of sum-rule results which establish precise connection between the moments of $G(0,0;{\bf q})$ and thermodynamic singularities \cite{pe2,ep}. The most important of these are
\eqan{sumrules}
G_0(0,0) &\sim & \tau^{-\alpha_s} \nonu \\
G_2(0,0) &\sim& \sigma_{\alpha \beta} \tau^{-2(1+\beta_s)}
\naqe
and recall that $\alpha_s=\beta_s=0$ in MF theory. In addition we note that \cite{ep}
\equ
\int_0^\infty dz G_0(0,z) = \frac{\partial m_1}{\partial h} \sim \tau^{-(1+\beta_s)}
\uqe
which shows the role played by correlations from the interface to the wall.    

A more precise MF expression for $G^{w\beta}(0,0;{\bf q})$ at critical wetting will be derived later using the generalized stiffness matrix formalism. The point that we wish to emphasise here is that the CW and FJ models do not satisfy the exact sum-rules in a manner that is also consistent with the non-Lorentzian form of $G(0,0;{\bf q})$ seen in the MF LGW calculations. The reasons for this will become clearer in the next few sections.
\end{list}

\subsection{Outline}     
In this paper we further develop the idea forwarded by PB that the problems of wetting theory reflect the failure of effective Hamiltonian models based on a single collective coordinate to account for the coupling of interfacial fluctuations. Thus the physical mechanism underlying our mathematical modelling is the following {\bf coupling hypothesis} 
\begin{quote}
The correlation function near the wall and some of the non-universal physics associated with wetting at the upper critical dimension $d=3$ is sensitive to the coupling of order parameter fluctuations near the unbinding interface and wall.
\end{quote}
PB concentrated on the complete wetting transition and constructed a two-field model $H[\ell_1,\ell_2]$ which in Gaussian approximation (generalising the calculations leading to (\ref{mfident})) identically recovered the MF correlation functions near the interface {\it and} wall. Here the generalized collective coordinates $\ell_1$ and $\ell_2$ represent surfaces of fixed magnetization which remain bound and unbind from the wall respectively. Subsequent RG \cite{bp1,bp,bp2} analysis of the two-field model for fluctuation effects in $d=3$ showed that the effective value of the wetting parameter determining the critical amplitude $\theta$ was renormalized compared to the CW expression. Thus addressing (P3) for complete wetting appears to resolve the second problem (P2), at least semi-quantitatively. Unfortunately the PB model is not well suited to studying critical wetting and only hints at a possible influence of coupling on this transition \cite{pb3}. 

The task of the present work is to generalise the PB approach to obtain a quantitative theory of coupling effects at complete and critical wetting. To do this requires a more thorough analysis of the connection between microscopic models (such as the LGW Hamiltonian) and collective coordinate theories. As we shall show it is first necessary to broaden the generalized effective Hamiltonian approach to allow for different types of collective coordinates describing order parameter fluctuations near the wall. Having done this we argue that it is possible to chose an optimal definition using a novel variational principle such that the sum over all possible configurations appearing in the effective theory most accurately mimics the partition function for the microscopic model. The optimal Hamiltonian, written $H[s,\ell]$, is then studied to elucidate the influence of coupling on the surface phase diagram and critical properties.

Our paper is arranged as follows: we begin by reviewing the PB theory and its successes in addressing (P2) and (P3) for complete wetting. The breakdown of the PB description near critical wetting is then discussed in addition to some pointers for possible coupling effects at this transition. In Sec.\ \ref{generalized} we introduce further examples of coupled Hamiltonians which use different types of collective coordinate at the wall. We provide a proof of a generalized correlation function reconstruction scheme (CFRS) which shows that the physical correlation function  $G(z_1,z_2;{\bf q})$ may be recovered (at MF level) using arbitrary choices of the collective coordinates, provided they satisfy certain conditions of locality. The invariance of $G(z_1,z_2;{\bf q})$ is crucial to our further analysis. 

In Sec.\ \ref{optimal} we introduce a particular class or set ${\cal H}$ of coupled models. This smoothly interpolates between two cases which describe order parameter fluctuations near the wall as local translations and enhancements of the magnetization respectively. The set ${\cal H}$ contains more general types of collective coordinate (which we shall refer to as proper coordinates) which have both interfacial- and spin-like components. All of the models in ${\cal H}$ are candidates for describing coupling effects beyond MF theory. We forward a novel optimization scheme leading to a unique choice of proper collective coordinate model in ${\cal H}$ which best describes fluctuations at the wall coupled to the unbinding interface. 

In Sec.\ \ref{theop} we consider fluctuation effects and show that the extent of the coupling is controlled by an angle-like variable $\delta^*$ whose value depends on the proximity of the critical wetting phase boundary. In particular for $T \gg T_W$ deep in the complete wetting regime the angle $\delta^* \approx \frac{\pi}{2}$ and the optimal theory reduces to the PB model. As the temperature is reduced, so that the thermodynamic path (iii) is closer and closer to $T_W$, the angle $\delta^*$ rotates to zero signifying a qualitative change in the role of coupling. This mechanism yields a precise expression for the temperature dependence of the renormalized wetting parameter determining $\theta$ in the proximity of $T_W$. For critical wetting the coupling has the effect of dramatically decreasing the extent of the true asymptotic critical regime for the local susceptibility $\chi_1$ consistent with the Ising model MC observations of MF-like behaviour (problem (P1)). The calculations for both critical and complete wetting highlight the role played by a second wetting parameter (written $\Omega$) associated with fluctuations near the wall which has no counterpart in CW and FJ theories. We discuss the temperature dependence of the new wetting parameter and argue that it approaches a universal value $\Omega_C$ close to unity as $T \rightarrow T_C^-$.      

The justification for the coupling hypothesis is largely {\it a posteriori} and relies on the success/failure of the theory to explain problems (P1)--(P3). Nevertheless we conclude our article with some remarks suggesting that the coupling to order parameter fluctuations at the wall can only effect thermodynamic singularities at the upper critical dimension for systems with short-range forces.

\section{Stiffness matrix formalism for complete wetting}
\label{stiffnessmatrix}

We begin our discussion of coupled fluctuations by reviewing the formalism and results of the PB theory of complete wetting which will be generalized in subsequent sections. To this end we first summarise the main results establishing the connection between FJ theory and MF correlation functions (further details may be found in \cite{pb,pa}).

\subsection{FJ theory and correlation functions} 
\label{FJtheory} 
The systematic structure of the FJ theory allows precise connection to be made with the MF correlation function of the LGW theory satisfying the Ornstein-Zernike equation (\ref{ozdif}). Following Parry \cite{pa}, consider the continuous set of FJ Hamiltonians $\{ H_{FJ}[\ell;m^X] \}$ by allowing for all possible values of $m^X$ belonging to the range of magnetizations $\tilde{m_1} \ge m^X > m_\beta$ seen in the MF profile $\tilde{m}(z)$. Ignoring fluctuations, the equilibrium position $\tilde{\ell} \equiv z$ of the surface of fixed magnetization $m^X$ clearly satisfies $m^X = \tilde{m}(z)$. Now consider the corresponding structure factor
\equ
S({\bf q};z) \equiv \int d{\bf y}_{12} {\rm e}^{i {\bf q}.{\bf y}_{12}} \langle (\ell({\bf y}_1) - z)(\ell({\bf y}_2)-z) \rangle
\uqe
In Gaussian approximation, valid for small fluctuations, this is trivially calculated
\equn{2S}
S({\bf q};z) = \frac{1}{W'' \Bigl( z;\tilde{m}(z) \Bigr) + q^2 \Sigma \Bigl( z;\tilde{m}(z) \Bigr)}
\nuqe
where
\equn{wdif}
W'' \Bigl( z;\tilde{m}(z) \Bigr) \equiv \frac{d^2}{d \ell^2} W(\ell;m^X)
\nuqe
for $\ell=z$ and $m^X = \tilde{m}(z)$. Note that in writing (\ref{2S}) we have ignored terms of order $q^4$ that would arise if we included an expression for the rigidity in the FJ model. However these are not important for interfacial phenomena and will not be discussed further (see \cite{pb4} for explicit calculations). It transpires that there is a remarkably elegant relation between the set $\{ S({\bf q};z) \}$, parameterized by $z$, and the MF correlation function  $G(z,z;{\bf q})$. To see this note that within FJ theory a small local translation (say) in the position of the surface of fixed magnetization $m^X$ generates a precise change in the magnetization
\equ
\delta m({\bf y}) = \frac{\partial m_\pi}{\partial \ell_\pi} (z;\ell_\pi) \delta \ell({\bf y})
\uqe 
at position $z$. Therefore a Gaussian approximation for each of the FJ Hamiltonians in the set may be used to calculate an expression for a magnetization pair correlation function at any two points. Let us denote the result for the Hamiltonian $H[\ell;m^X]$ by $G^{FJ}(z_1,z_2;{\bf q};m^X)$. Then we can identify
\equ
G^{FJ}(z_1,z_2;{\bf q};m^X) = \frac{\partial m_\pi}{\partial \ell_\pi}(z_1;\ell_\pi) \frac{\partial m_\pi}{\partial \ell_\pi}(z_2;\ell_\pi) S({\bf q};z)
\uqe
where the values of $m^X$ and $z$ on the l.h.s.\ and r.h.s.\ are related by $m^X=\tilde{m}(z)$ and the partial derivatives of the planar constrained profile are evaluated at equilibrium $\ell_\pi=\tilde{\ell} = z$. The question now remains how does this set of correlation functions relate to the actual solution $G(z_1,z_2;{\bf q})$ of the MF Ornstein-Zernike equation (\ref{ozdif})? Explicit calculation shows that each element $H_{FJ}[\ell;m^X]$ generates the correct MF correlation function only for positions $z_1=z_2=z$. Specifically  we can identify
\equ
G(z,z;{\bf q}) = G^{FJ}(z,z;{\bf q};\tilde{m}(z))
\uqe
which becomes
\equn{cfrs1}
G(z,z;{\bf q}) =\tilde{m}'(z)^2 S({\bf q};z)
\nuqe
on using
\equ
\frac{\partial m_\pi}{\partial \ell_\pi}(\tilde{\ell};\ell_\pi) = - \tilde{m}'(\tilde{\ell})
\uqe                                                         
(obtained by simple differentiation of the crossing criterion (\ref{CC})). For example, explicit evaluation of (\ref{wdif}) in FJ theory yields
\eqa
G_0(z,z) &=& \tilde{m}'(z)^2 S(0;z) \nonu \\
&=& \tilde{m}'(z)^2 \left( \frac{1}{\tilde{m}'_1 (c \tilde{m}'_1 - \tilde{m}''_1)} + \int_0^z \frac{dz'}{\tilde{m}'(z')^2} \right)
\aqe
which is the analytic solution to (\ref{ozdif}). These remarks make it clear why the FJ model with $m^X=0$ (and hence the CW model) fail to describe the wall correlation function $G(0,0;{\bf q})$ since the choice of $m^X$ is entirely inappropriate. With $m^X=0$ the FJ model accurately describes fluctuations near the $\alpha \beta$ interface and can exactly recover the MF $G(z,z;{\bf q})$ only at the position $z$ where $\tilde{m}(z)=0$.

\subsection{Two-field theory and correlation functions}
\label{twofield}
\begin{figure}[h]
\begin{center}
\scalebox{0.5}{\includegraphics{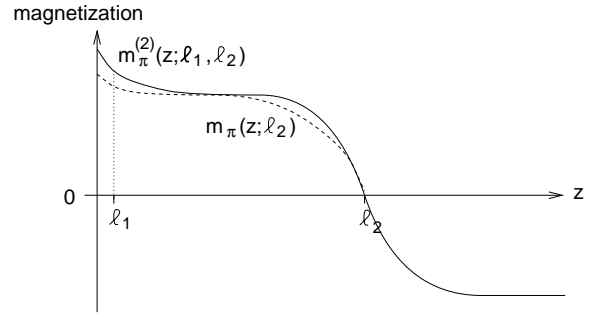}}
\end{center}
\caption{Schematic portrayal of the single (dashed line) and doubly constrained planar profiles near a complete wetting transition. Discontinuities in the gradients exist at $z=\ell_1$ and $z=\ell_2$ but these have not been explicitly shown.}
\label{bp}
\end{figure}

PB next consider the properties of two-field models $H[\ell_1,\ell_2;m^X_1,m^X_2]$ for surfaces of fixed magnetization $m^X_1$ and $m^X_2$ described by a pair of collective coordinates $\ell_1$ and $\ell_2$ \cite{pb}. Importantly for the complete wetting transition, on which they concentrate, the MF and FJ profile has a significant lip at the wall (see Fig.\ \ref{bp}) so that it is possible to chose $\tilde{m}_1 \gtrsim m^X_1 > m_\beta$ and $m^X_2=0$ (say) so that
\setcounter{c}{0}
\begin{list}{(\roman{c})}{\usecounter{c} }
\item the upper surface unbinds 
\item the lower surface remains close to the wall
\end{list}
as $\bar{h} \rightarrow 0$. 
Consequently the MF positions $\tilde{\ell_\mu} = z_\mu$ with $\mu=1,2$ of these generalized surfaces represent planes that are close to the wall and unbinding interface. The two-field Hamiltonian is constructed using a natural generalization of the FJ method. The fundamental saddle point identification is
\equ
H[\ell_1,\ell_2] = H_{LGW}[m_\Xi^{(2)}({\bf r};\ell_1,\ell_2)]
\uqe
where we have dropped the explicit $m^X_\mu$ dependence. Here $m_\Xi^{(2)}$ represents the magnetization distribution which minimises the LGW Hamiltonian subject to the double crossing criterion
\equ
m_\Xi^{(2)}( {\bf r}; \ell_1,\ell_2) = m^X_\mu \mbox{ \hspace*{1cm} with ${\bf r}= ({\bf y},\ell_\mu)$}
\uqe
As mentioned earlier calculations are eased when we observe that only the properties of planar constrained profiles $m_\pi^{(2)}(z;\ell_1,\ell_2)$ need be considered to derive models with gradient terms. 

In this way PB find \cite{pb}
\equ
H[\ell_1,\ell_2] = \int d{\bf y} \left\{ \frac{1}{2} \Sigma_{\mu \nu}(\ell_1,\ell_2) \nabla \ell_\mu . \nabla \ell_\nu + W_2(\ell_1,\ell_2) \right\}
\uqe
where the binding potential is 
\eqan{GEN1}
\lefteqn{W_2(\ell_1,\ell_2) = \Biggl\{ \phi_1 \Bigl( m_{\pi1}^{(2)} \Bigr)} \nonu \\
& & \left.  + \int_0^\infty dz \left[ \frac{1}{2} \left( \frac{\partial m_\pi^{(2)}}{\partial z} \right)^2 + \phi(m_\pi^{(2)}) \right] \Biggr\} \right|_{\ell_{i \pi}=\ell_i({\bf y})}
\naqe
ignoring some constant terms independent of $\ell_2$ (which are subtracted off). The stiffness matrix elements are given by
\equn{GEN2}
\Sigma_{\mu \nu}(\ell_1,\ell_2) = \left. \int_0^\infty dz \frac{\partial m_\pi^{(2)}}{\partial \ell_{\mu \pi}} . \frac{\partial m_\pi^{(2)}}{\partial \ell_{\nu \pi}} \right|_{\ell_{i \pi}=\ell_i({\bf y})}
\nuqe

The constrained profile satisfies the same Euler-Lagrange (\ref{el}) considered earlier but must now be solved subject to the double planar crossing criterion. Once $m_\pi^{(2)}(z;\ell_1,\ell_2)$ is found then the explicit position dependence of   $W_2(\ell_1,\ell_2)$ and $\Sigma_{\mu \nu}(\ell_1,\ell_2)$ may be calculated. Conveniently the binding potential decomposes into a shielded form
\equ
W_2(\ell_1,\ell_2) = U(\ell_1) + W(\ell_2-\ell_1)
\uqe
where $W(\ell)$ is similar to the CW result (\ref{W}). The first term simply serves to bind the lower surface to the wall. We will always assume that the fluctuations $\xi_\perp^{(1)} \equiv \Bigl[ \langle \delta \ell_1^2 \rangle \Bigr]^{-1/2}$ are small so that we may write for complete wetting
\equ
W_2(\ell_1,\ell_2) = \frac{r}{2} \ell_1^2 + \bar{h} \ell_{21} + 2 \kappa m_\alpha \tau {\rm e}^{-\kappa \ell_{21}} + \cdots
\uqe
with $r \sim \tau^2$ explicitly determined and $\ell_{21}=\ell_2-\ell_1$. Minimization of $W_2(\ell_1,\ell_2)$ gives the MF positions $z_1,z_2$ of the collective coordinates. The position dependence of the stiffness coefficients can also be determined 
\eqan{s12bp}
\Sigma_{11} &\sim& \sigma_{w\alpha} - \phi_1(\tilde{m}_1) + O({\rm e}^{-\kappa \ell_{21}}) \nonu \\
\Sigma_{22} &\sim& \sigma_{\alpha\beta} + O({\rm e}^{-\kappa \ell_{21}}) \nonu \\
\Sigma_{12} &\sim& \kappa^2 m_\alpha \tau \ell_{21} {\rm e}^{-\kappa \ell_{21}}
\naqe
and notice that $\sigma_{w\alpha} - \phi_1(\tilde{m}_1) \sim \tau^2$. In these expressions $\tilde{m}_1$ is to be interpreted as the magnetization $m^X_1$ for the wall-$\alpha$ interface. In fact the position dependence of the stiffness coefficients plays no part in determining leading order thermodynamic effects although the cross term $\Sigma_{12}$  is important as regards the coherent and next-to-leading order correlation function behaviour discussed earlier \cite{pb,pb1}.     

Due to the inclusion of two collective coordinates local to the $\alpha \beta$ interface and wall, the coupled model can fully explain the non-Lorentzian structure of $G(0,0;{\bf q})$ described in (P3). In fact precise connection with MF correlations can be made using the following CFRS \cite{pb1}. Define the structure factor matrix ${\bf S}({\bf q})$ as the $2 \times 2$ matrix with elements
\equn{Sdef}
S_{\mu \nu}({\bf q}) = \int d{\bf y}_{12} {\rm e}^{i {\bf q}.{\bf y}_{12}} \langle \delta \ell_\mu({\bf y}_1) \delta \ell_\nu({\bf y}_2) \rangle
\nuqe
These can easily be calculated using the relation
\equn{sdef}
{\bf S}^{-1}({\bf q}) = \left( \begin{array}{ll} \partial^2_{11} & \partial^2_{12} \\ \partial^2_{21} & \partial^2_{22} \end{array} \right) W(\ell_1,\ell_2) + q^2 {\bf \Sigma}
\nuqe
where $\partial^2_{\mu \nu} \equiv \frac{\partial^2}{\partial \ell_\mu \partial \ell_\nu}$ and is evaluated at equilibrium. From the matrix elements one can calculate the MF expressions for three possible correlation functions by
\equn{gsr}
G(z_\mu,z_\nu;{\bf q}) = \tilde{m}'(z_\mu) \tilde{m}'(z_\nu) S_{\mu \nu}({\bf q})
\nuqe
where $z_\mu,z_\nu \in \{ z_1,z_2 \}$, i.e.\ $m^X_\mu=\tilde{m}(z_\mu)$. This identically recovers the known solution to the Ornstein-Zernike equation (\ref{ozdif}). 

For example, at the wall PB derive 
\equn{nozg}
G(0,0;{\bf q}) = \frac{\tilde{m}_1^{'2}}{r + q^2 \left[ \Sigma_{11} + \frac{\Sigma_{22} + 2 \Sigma_{12}}{1+q^2 \xi_\parallel^2} \right] + O(q^4)}
\nuqe
which is of the required form (\ref{PBresult}). PB proceed to show that the elements of the stiffness matrix obey the relation \cite{pb,pb1}
\equn{smfe1}
\sum_{\mu,\nu} \Sigma_{\mu \nu}(0,z_2) = \sigma_{w\alpha}-\phi_1(\tilde{m}_1)
\nuqe
which ensures that the sum-rule (\ref{sumr}) describing the coherence of long wavelength fluctuations is exactly satisfied. The origin of the $f_{\rm sing}$  contribution to $G_2(0,0)$ is elegantly explained by this approach through the position dependence of the off-diagonal stiffness matrix term since
\equn{smfe2}
2\Sigma_{12}(0,z_2) \approx f_{\rm sing}
\nuqe
and recall that the arguments refer to the mean positions of $\ell_1$ and $\ell_2$. Notice that the position dependence of $\Sigma_{12}$ (see (\ref{s12bp})) is longer ranged than $\Sigma_{11}$ and $\Sigma_{22}$ and is precisely of the form required to yield the singularity $f_{\rm sing} \sim \bar{h} \ln |\bar{h}|$. Following PB we refer to (\ref{smfe1}) and (\ref{smfe2}) as stiffness matrix free energy relations.

\subsection{Renormalization of the wetting parameter}  
\label{renorma} 
The most important consequence of coupling is independent of the position dependence of the stiffness matrix elements and it suffices to write the Hamiltonian \cite{bp}
\equ
H[\ell_1,\ell_2] = H_0[\ell_1,\ell_2] + H_1[\ell_1,\ell_2]
\uqe
where  
\equ
H_0[\ell_1,\ell_2] = \int d{\bf y} \left\{ \frac{1}{2} \Sigma_{11} (\nabla \ell_1)^2 + \frac{1}{2} \Sigma_{22} (\nabla \ell_2)^2 + \frac{1}{2} r \ell_1^2 \right\}
\uqe                                                     
and 
\equ
H_1[\ell_1,\ell_2] = \int d{\bf y} W(\ell_2,\ell_1)
\uqe                                                      

In this approximation we may identify $\Sigma_{22} = \Sigma_{\alpha \beta}$ and $\Sigma_{11} = \sigma_{w\alpha} - \phi_1(\tilde{m}_1)$. Note that $\sqrt{\frac{\Sigma_{11}}{r}} \equiv \xi_{w\alpha}$ corresponds to the {\it finite} transverse (second-moment) correlation length for surface correlations at the wall-$\alpha$ interface. This can be reliably calculated in MF approximation away from $T_C$ yielding $\xi_{w\alpha} = [\kappa(c+\kappa) ]^{-1/2}$. The momentum cut-off $\Lambda_2$ for the upper field is the same as that for the CW and FJ models (denoted $\Lambda$). Similarly the momentum cut-off $\Lambda_1$ for the lower field must be chosen appropriately such that the r.m.s.\ fluctuations are small (and typically of order a bulk correlation length) which requires $\Lambda_1 \ll \sqrt{\Sigma_{11}}$ \cite{pbs}. 

As the fluctuations of $\ell_1$ are adequately described in Gaussian approximation it is possible to account for their renormalization exactly. The interaction $H_1[\ell_1,\ell_2]$ is then treated in linear approximation analogous to the RG theory of Fisher and Huse \cite{fh} and FJ \cite{jinf1}. First the fast modes (appearing in the Fourier representations of the fields) with wave-vectors $q$ satisfying $\Lambda_\mu/b < q < \Lambda_\mu$ are integrated out and the Hamiltonian rescaled leaving $H_0[\ell_1,\ell_2]$ invariant. The flow equation for $W^{(t)}(\ell_1,\ell_2)$ reads
\equn{RGdif}
\left[ 2 - \frac{\partial}{\partial t} + \omega_1 \frac{\partial^2}{\partial \ell_1^2} + \omega_2 \frac{\partial^2}{\partial \ell_2^2}  \right] W^{(t)}(\ell_1,\ell_2) = 0 
\nuqe
with $t=\ln b$ the infinitesimal rescaling factor and
\eqa
\omega_1 &=& \frac{\kappa^2}{4 \pi \Sigma_{11}} \left( 1- \frac{r}{\Sigma_{11} \Lambda_1^2 + r} \right) \\
\omega_2 &=& \frac{\kappa^2}{4\pi \Sigma_{\alpha \beta}}
\aqe

The diffusion equation (\ref{RGdif}) can be easily solved
\eqan{RGr}
W^{(t)}(\ell_1,\ell_2) &=& \frac{{\rm e}^{2t}}{4\pi t\sqrt{\omega_1 \omega_2}} \int d\ell'_1 d\ell'_2 W^{(0)}(\ell_1',\ell_2') \nonu \\
& & \times \exp \left( - \frac{(\ell_1-\ell_1')^2}{4\omega_1 t} - \frac{(\ell_2-\ell_2')^2}{4\omega_2 t} \right) 
\naqe
where $W^{(0)}(\ell_1,\ell_2)$ is the bare interaction contribution to the binding potential. In the PB model the fields only interact through a relative term which is a function of $\ell_{21} = \ell_2-\ell_1$ so that (\ref{RGr}) simplifies to
\equ
W^{(t)}(\ell_{21})=\frac{{\rm e}^{2t}} {\sqrt{4\pi(\omega_1+\omega_2)t}} \int_{-\infty}^\infty d\ell'  W^{(0)}(\ell') {\rm e}^{-\frac{(\ell_{21}-\ell')^2}{4(\omega_1+\omega_2)t}}
\uqe
which needs to be studied at the matching point $t^*$ \cite{fh} where the curvature $W^{(t^*)''}(\langle \ell_{21} \rangle)=\Sigma_{\alpha \beta} \kappa^2$. At this point we may identify $ e^{t^*}$ as $\kappa \xi_\parallel$. These remarks establish that the binding potential renormalizes exactly as in the CW theory but the divergence of $\langle \ell \rangle$ is determined by a renormalized wetting parameter 
\eqan{barom}
\bar{\omega} &=& \omega_1+\omega_2 \nonu \\
&=& \omega + \frac{\kappa^2}{4 \pi \Sigma_{11}} \left( 1 - \frac{r}{\Sigma_{11} \Lambda_1^2 +r} \right) 
\naqe                   
The value of $\bar{\omega}$ lies between two extremes corresponding to $r=\infty$ and $r= 0$ representing cases where the $\ell_1$ fluctuations are suppressed and free respectively. 

The implication of the PB model is that the adsorption critical amplitude $\theta$ is increased due to coupling effects
\equn{thetaR}
\theta = 1 + \frac{1}{2} \left( \omega + \frac{\kappa^2}{4 \pi \Sigma_{11} [1+(\Lambda_1 \xi_{w\alpha} )^{-2}]} \right)
\nuqe
which should be compared to the CW result (\ref{thetacw}) (assuming that $\omega,\bar{\omega}<2$). PB estimate \cite{bp} that the increment $\Delta \theta$ to the CW result is about $0.3$ for Ising-like systems deep in the complete wetting regime yielding a value for $\theta \approx 1.7$  close to the BLF simulation result.     
Importantly the linear RG expression (\ref{thetaR}) is supported by non-linear RG analysis \cite{bp2} and also by MC simulation of a discretized version of the model \cite{sf}.

\subsection{Interpretation}
\label{interp}
Before we discuss the unsatisfactory features of the PB model we make some new remarks concerning the interpretation of the fundamental results (\ref{barom}) and (\ref{thetaR}). The first of these is that the renormalization of the critical amplitude $\theta$ can be understood heuristically using a simple generalization of the elegant scaling analysis of Lipowsky and Fisher \cite{lf} which takes into account the asymptotic coherence of long wavelength fluctuations at complete wetting.

To see this we initially recover the standard result (\ref{thetacw}) using scaling arguments based on the CW model (\ref{inter}). Neglecting fluctuation effects the singular contribution to the free energy per unit area $A$ of the wall is clearly
\eqan{MFf}
f_{\rm sing}^{MF} &=& W(\tilde{\ell}) \nonu \\
&=& \frac{\bar{h} \ln |\bar{h}|^{-1}}{\kappa} + O(\bar{h})
\naqe
corresponding to MF theory. Now following Lipowsky and Fisher we suppose that the fluctuations in the position of the $\alpha \beta$ interface lead to an extra bending energy so that
\equ
f_{\rm sing}^{CW} \approx \half \Sigma_{\alpha \beta} \left( \frac{\xi_\perp}{\xi_\parallel} \right)^2 + f_{\rm sing}^{MF}
\uqe
Using the known interfacial roughness (\ref{gas}) and the divergence of $\xi_\parallel$ (which is unchanged from the MF result (\ref{mf2})) we are led to
\equ
f_{\rm sing}^{CW} \approx (1+\frac{\omega}{2}) \frac{\bar{h} \ln |\bar{h}|^{-1}}{\kappa}
\uqe 
which is the explicit RG result (for $\omega<2$) consistent with (\ref{thetacw}) (recall that $\langle \ell \rangle = \frac{\partial f_{\rm sing}}{\partial \bar{h}}$).

To include coupling effects we first assume (as is reasonable) that the estimate of the bending energy contribution from the large scale capillary-wave-like fluctuations of the $\ell_2({\bf y})$ field is unchanged i.e.\ the roughness relation between $\xi_\perp$ and $\xi_\parallel$ of the $\alpha \beta$ interface is the same as the CW model (see below). Further, we suppose that the only effect of the small fluctuations in the $\ell_1({\bf y})$ field is to increase the effective area of the wall as seen by asymptotically long wavelength (coherent) fluctuations in the $\alpha \beta$ interface. The relative increase in area can be estimated (analogous to the bending energy contribution) as
\equ 
\frac{A'}{A} = 1 + \half \Bigl( \kappa \xi_\perp^{(1)} \Bigr)^2
\uqe
where
\equ
\Bigl( \kappa \xi_\perp^{(1)} \Bigr)^2 = \frac{k_BT \kappa^2}{4\pi \Sigma_{11}} \ln \Bigl( 1 + (\Lambda_1 \xi_{w\alpha})^2 \Bigr)
\uqe
determines the r.m.s.\ fluctuation of the lower surface. Now due to the asymptotic coherence of long wavelength fluctuations the direct MF contribution to the singular free energy must be modified due to the increase in the effective area of the wall. Thus we must add a coupling contribution 
\equ
 \half \Bigl( \kappa \xi_\perp^{(1)} \Bigr)^2 W(\tilde{\ell})
\uqe
which with the usual MF bending energy term gives
\equ
f_{\rm sing}^{PB} \approx \left( 1 + \frac{\omega}{2} + \frac{k_BT \kappa^2}{8\pi \Sigma_{11}} (\Lambda_1 \xi_{w\alpha})^2 \right) \frac{\bar{h} \ln |\bar{h}|^{-1}}{\kappa} 
\uqe
and correctly identifies the renormalized expansion for $\theta$ for small $\Lambda_1$. 

A nice feature of this argument is that it directly relates the renormalization of $\theta$ to the coherence of asymptotic long wavelength fluctuations signifying the importance of similar coupling effects for problems (P1) and (P3).

Implicit in the above is the assumption that the roughness of the $\alpha \beta $ interface is not changed by coupling effects. This takes us to our second remark that statements concerning the renormalization of the wetting parameter only refer to the critical amplitude $\theta$ and not the value of $\omega$ determining the divergence of the roughness $\xi_\perp$. In fact this can be easily deduced from the renormalized CFRS which identifies the structure factors $S_{\mu \nu}({\bf q})$ (defined in (\ref{Sdef})) beyond MF approximation in $d=3$ \cite{pb3}. 

The renormalized CFRS is very similar to that described earlier because the RG matching procedure reduces the problem to a Gaussian calculation just as in MF theory. Thus we need only replace the binding potential and stiffness matrix by their renormalized values and allow for the spatial rescaling factor. We first calculate the renormalized (inverse) matrix
\equ
\Bigl( {\bf S}^{(t^*)}({\bf q}) \Bigr)^{-1} = \left( \begin{array}{ll} \partial^2_{11} & \partial^2_{12} \\ \partial^2_{21} & \partial^2_{22} \end{array} \right) W^{(t^*)}_2(\ell_1,\ell_2) + q^2 {\bf \Sigma} {\rm e}^{2 t^*}
\uqe
where 
\eqa
W_2^{(t)} &\equiv& {\rm e}^{2 t} \half r \ell_1^2 + W^{(t)}(\ell_{21}) \nonu \\
& = & {\rm e}^{2t} \left( \half r \ell_1^2 + \bar{h} \ell_{21} + 2 m_\alpha \kappa \tau {\rm e}^{\bar{\omega}t-\kappa \ell_{21}}+ \cdots \right)
\aqe 
and we have assumed that $\bar{\omega}<2$. Ignoring the position dependent elements, the stiffness matrix is
\equ
{\bf \Sigma} = \left( \begin{array}{ll} \Sigma_{11} & 0 \\ 0 & \Sigma_{\alpha \beta} \end{array} \right) 
\uqe
and is not renormalized (analogous to the stiffness coefficient in CW theory \cite{lf}). The required structure factors then follow as
\equ
S_{\mu \nu}(q) = {\rm e}^{2 t^*} S_{\mu \nu}^{(t^*)}(q)
\uqe
which are easily calculated and yield final results that are basically unchanged from the MF calculations described earlier. In particular the $S_{22}(q)$ element determines the mean square value of the Fourier amplitude $\delta \hat{\ell}_2({\bf q})$ (see \cite{pb1})
\eqa
S_{22}(q) &=& \langle | \delta \hat{\ell}_2({\bf q}) |^2 \rangle \nonu \\
&=& \frac{{\rm e}^{2t^*}}{{\rm e}^{2t^*} \Sigma_{\alpha \beta} q^2 + \frac{W^{(t^*)''}(r+\Sigma_{11} q^2) {\rm e}^{2t^*}}{W^{(t^*)''}+{\rm e}^{2t^*} (r + q^2 \Sigma_{11})}} \nonu \\
&\approx & \frac{ {\rm e}^{2 t^*}}{W^{(t^*)''}+ \Sigma_{\alpha \beta} q^2 {\rm e}^{2t^*}} \nonu \\
&=& \frac{1}{\Sigma_{\alpha \beta}(\xi_\parallel^{-2}+q^2)}
\aqe
using $W^{(t^*)''}=\Sigma_{\alpha \beta} \kappa^2$ and ${\rm e}^{t^*}=\kappa \xi_\parallel$. Performing the Fourier inversion in the standard fashion yields
\equn{tr}
\Bigl( \kappa \xi_\perp \Bigr)^2 \approx \omega \ln \left( 1 + (\Lambda \xi_\parallel)^2 \right)
\nuqe
as quoted in the introduction for the CW model. Thus the effective value of the wetting parameter is unaltered as regards the interfacial roughness relation in contrast to the calculation involving the critical amplitude $\theta$. 

Combining (\ref{tr}) with the two-field prediction for the transverse correlation length in the thin-film geometry with opposite surface fields (\ref{oppos}) yields
\equn{width}
\xi_\perp = \sqrt{\frac{\omega D}{2\theta \kappa}} \mbox{\hspace*{9mm} for $D \goto \infty$}
\nuqe
in the soft mode regime. This shows that the divergence of the roughness as $D \goto \infty$ is controlled by the bare {\it and} renormalized wetting parameters (see (\ref{thetaR})). Data for the mid-point width $\frac{m_\alpha-m_\beta}{m'(D/2)}$ (denoted as $w=\sqrt{2\pi} \xi_\perp$ in \cite{kerle}) have been extracted from the BLF simulation data \cite{kerle} at $\frac{T}{T_C} = 0.9554$ and compare favourably with the asymptotic prediction above for moderately wide thin films $D \le 40$. Note that the prediction (\ref{width}) correctly identifies the pre-factors for the square root divergence of $\xi_\perp$ (and hence $w$) misquoted in \cite{kerle} although the numerical estimates for $w$ are very similar to their Fig.\ 3(b).

Also note that the present RG result for $S_{11}(q)$ gives an identical non-Lorentzian expression as that found in the MF calculation (\ref{PBresult}) strongly suggesting that the exact $G(0,0;{\bf q})$ has such a structure.

\subsection{Problems with the description of critical wetting}
\label{problemswith}

\subsubsection{Crossover from complete to critical wetting}      
The PB model is certainly not capable of yielding a fully quantitative theory for the temperature dependence of the renormalized critical amplitude $\theta$ which could be compared in detail with the simulation results. This is due to the uncertainty in the value of the momentum cut-off $\Lambda_1$ for the lower field which appears in the RG prediction. This is a serious defect of the model which forces us to rethink the nature of order parameter fluctuations near the wall in the vicinity of $T_W$. To see this we follow PBS and consider complete wetting transitions occurring along paths (iii) in Fig.\ \ref{pd} closer and closer to the wetting temperature $T_W$. Recalling that the stiffness coefficient for the lower surface vanishes like $\Sigma_{11} \sim \tau^2$ we are forced to conclude that the cut-off $\Lambda_1$ must also vanish in this limit otherwise the value of $\bar{\omega}$ is singular (see (\ref{thetaR})). In fact such behaviour can be anticipated since as mentioned earlier we need to impose the inequality $\Lambda_1 \ll \sqrt{\Sigma_{11}}$ in order that the r.m.s.\ amplitude of the Gaussian fluctuations $\ell_1({\bf y})$ at the wall remains bounded. This pathology is intimately linked to the behaviour of the MF and FJ profiles near the wall which flatten as the critical wetting transition temperature is approached, signalling the breakdown of the crossing criterion method used by PB to model small order parameter fluctuations. In fact this feature is also seen in the equilibrium magnetization profiles extracted from the Ising model simulation studies --- see for example Fig.\ 4 of \cite{blf1}. PBS point out that this has important consequences for complete wetting close to $T_W$ since it implies that the increment to $\bar{\omega}$ due to the coupling of fluctuations must vanish
\equn{wlimit}
\stackrel{\lim}{\scs T\rightarrow T_W^+} \Bigl\{ \bar{\omega} \Bigr\} = \omega(T_W)
\nuqe
so that
\equ
\stackrel{\lim}{\scs T\rightarrow T_W^+} \Bigl\{ \theta \Bigr\} = 1+\frac{\omega}{2}
\uqe
This prediction is in very good agreement with the BLF simulation data (see Fig.\ \ref{bigtheta}) and extrapolation to $T_W \approx 0.9T_C$ yields $\theta \approx 1.4$ and hence $\omega(T_W) \approx 0.8$ which is close to the series expansion estimate of Fisher and Wen at this temperature \cite{fishw}. As conjectured by PBS the result (\ref{wlimit}) is indicative of some kind of decoupling phenomenon between order parameter fluctuations near the $\alpha \beta$ interface and wall occurring close to $T_W$. However the PB model can only describe this in the crudest possible (and unsatisfactory) way by allowing the cut-off $\Lambda_1$ to vanish as $T \rightarrow T_W^+$.

\subsubsection{Towards a theory of coupling at critical wetting}    
Despite the unwelcome behaviour of the cut-off $\Lambda_1$ during the approach to $T_W$ we note that the two-field theory does hint at a possible explanation of problem (P1). As pointed out by Parry and Boulter \cite{pb3} if we simply assume that the model, with a suitably chosen cut-off $\Lambda_1 \ll \tau^2$, is appropriate for describing coupled fluctuations along route (i) in Fig.\ \ref{pd} then it immediately follows from the RG transformations described earlier that the structure factors $S_{11}(0)$ and $S_{12}(0)$ associated with the lower field retain MF-like singularities even when $S_{22}(0)$ has a non-classical divergence with $\nu_\parallel$ given by (\ref{nu}). While we should be rightly suspicious of the precise status of this prediction one might tentatively suggest that coupling somehow reduces the manifestation of critical singularities for local response functions at the wall. As we shall see in the next sections this conjecture is supported by a more quantitative theory of coupling effects at wetting transitions.

\subsection{Open questions}    
While the semi-quantitative successes of the coupled model introduced by PB are encouraging there are a number of unpleasant features associated with the theory that warrant further investigation. All of these concern the pathological behaviour of the cut-off $\Lambda_1$ associated with the approach to the wetting temperature which clearly limit the application of the model. This is most unwelcome from a fundamental point of view since we should take a harsh opinion of any field theory (at least in condensed matter physics) in which the cut-off has singular behaviour.   

Our central goal in this paper is to develop a coupled theory of wetting transitions which involves a unified description of order parameter fluctuations at the wall. It should be equally applicable to complete and critical wetting and avoid the need to impose any special behaviour of the cut-off(s). The theory should also be able to explain the apparent success and simplicity of the PB model deep in the complete wetting regime and provide a more sensible mechanism for the decoupling of fluctuations associated with the result (\ref{wlimit}) (assuming that this survives).

\section{Generalized coupled models and the CFRS}
\label{generalized}    

To develop a fully quantitative model of coupling effects at wetting transitions it is necessary to reassess the role played by collective coordinates in effective Hamiltonian theory. The original motivation of PB was to construct an effective Hamiltonian which in Gaussian approximation could reproduce the Lorentzian and non-Lorentzian structure of the MF correlation function near the $\alpha \beta$ interface and wall respectively. As PB showed MF correlations could be precisely recovered using the CFRS described in the last section. We shall now that show that this is also possible for different choices of the collective coordinates. We begin with an illustrative example before considering the general case.

\subsection{Interfacial- and spin-like collective coordinates}     
\label{interfacialand}
In Sec.\ \ref{stiffnessmatrix} we reviewed how the two-field theory grew out of observations concerning the connection between MF correlations and the set of FJ Hamiltonians $\{ H[\ell;m^X] \}$. Heuristically, the crossing criterion naturally defines an interfacial-like collective coordinate $\ell({\bf y})$ which generates magnetization configurations by translating the position of a surface  of fixed magnetization. However it is also possible to define spin-like collective coordinates which model local enhancements of the magnetization in some plane. We begin by considering one-field models parallelling the analysis of Sec.\ \ref{FJtheory}.    

Let us focus on the behaviour of the $d-1$ dimensional plane of spins a distance $z$ from the wall and constrain them to satisfy
\equn{spinc}
m \Bigl({\bf y},z^X \Bigr) = \sigma({\bf y})
\nuqe
We may define an effective Hamiltonian for this spin plane by tracing over all magnetization configurations that leave $\{ \sigma({\bf y}) \}$ invariant
\equn{hsdef}
H[\sigma;z^X] = - \ln \left[ \int_C {\cal D} m \; {\rm e}^{-H_{LGW}[m]} \right]
\nuqe
where $C$ denotes the spin constraint (\ref{spinc}). Now let us assume, rightly or wrongly, that all fluctuations apart from those of the spin plane are small such that we can use a FJ-like saddle point identification
\eqa
H[\sigma;z^X] &=& \min_C H_{LGW}[m] \nonu \\
&=& H_{LGW}[m_\Xi]
\aqe
where $m_\Xi({\bf r};\sigma)$ is the profile that minimises $H_{LGW}[m]$ subject to the constraint (\ref{spinc}). Whether or not this assumption is justified will be considered later. For our present purpose of reconstructing MF correlations it is of no consequence as we now show.       

The constrained magnetization satisfies the usual Euler-Lagrange equation
\equn{2el}
-\nabla^2 m_\Xi + \phi'(m_\Xi) =0
\nuqe
with boundary conditions
\eqan{2bc}
\left. \frac{\partial m_\Xi}{\partial z} \right|_{z=0} &=& c m_{\Xi1}-h_1 \nonu \\
\stackrel{\rm lim}{\scs z \goto \infty} m_\Xi(z) &=& m_\beta
\naqe
and spin criterion (\ref{spinc}).

In the same manner as FJ theory however it suffices to consider planar constrained profiles $m_\pi(z;\sigma_\pi)$ from which $m_\Xi({\bf r};\sigma)$ may be constructed perturbatively using
\equ
m_\Xi({\bf r};\sigma) = m_\pi \Bigl( z;\sigma({\bf y}) \Bigr) + \frac{\partial m_\pi}{\partial \sigma_\pi} \delta \sigma({\bf y})
\uqe
with $\delta \sigma({\bf y})=\sigma({\bf y})-\sigma_\pi$. In this way it is straightforward to construct the effective spin plane Hamiltonian
\equn{HS}
H[\sigma;z^X] =  \int d{\bf y} \left\{ \frac{1}{2} \Sigma(\sigma;z^X) + W(\sigma;z^X) \right\}
\nuqe
(ignoring terms of $O(\nabla^2 \sigma)$) where
\eqa
\lefteqn{W(\sigma;z^X) = \Biggl\{ \phi_1(m_\pi(0;\sigma_\pi))} \nonu \\
& & \left.  + \int_0^\infty dz \left[ \half \left( \frac{\partial m_\pi}{\partial z} \right)^2 +\phi(m_\pi(z;\sigma_\pi)) \right]  \Biggr\} \right|_{\sigma_\pi=\sigma({\bf y})}
\aqe
and 
\equ
\Sigma(\sigma;z^X) =\left. \int_0^\infty dz \left( \frac{\partial m_\pi}{\partial \sigma_\pi} \right)^2 \right|_{\sigma_\pi=\sigma({\bf y})}
\uqe
By construction minimisation of $W(\sigma;z^X)$ recovers the MF value of the magnetization at position $z=z^X$. Thus we can identify
\equ
m_\pi(z;\tilde{\sigma}) = \tilde{m}(z)
\uqe
with
\equ
\tilde{\sigma} = \tilde{m}(z^X)
\uqe

So far our analysis parallels that of FJ but with a different type of collective coordinate. Now we follow Parry \cite{pa} and use the set of spin-plane Hamiltonians (by allowing all possible values of $z^X$) to derive a set of correlation functions $G^{(\sigma)}(z_1,z_2;{\bf q};z^X)$. Consider first the zeroth moments $G_0^{(\sigma)}(z_1,z_2;z^X)$. Each Hamiltonian can be used to calculate a correlation function since there is a one-to-one relation between fluctuations in $\sigma$ and the magnetization at any point ${\bf r}_1$. For homogeneous enhancements we note that
\equ
\delta m({\bf r}_1) = \frac{\partial m_\pi}{\partial \sigma_\pi}(z_1;\sigma_\pi) \delta \sigma({\bf y}_1) 
\uqe
and in particular at the plane $z=z^X$ itself
\equn{sigdif}
\frac{\partial m_\pi}{\partial \sigma}(z^X;\sigma)=1
\nuqe
Assuming that fluctuation are small and can be adequately described by a Gaussian approximation we arrive at the prediction
\equn{csig}
G_0^{(\sigma)}(z_1,z_2;z^X) = \left. \frac{ \frac{\partial m_\pi}{\partial \sigma_\pi}(z_1;\sigma_\pi) \frac{\partial m_\pi}{\partial \sigma_\pi}(z_2;\sigma_\pi)}{W''(\sigma;z^X)} \right|_{\tilde{\sigma}}
\nuqe
for the (zeroth) moment correlation function according to the model (\ref{HS}). Here
\equ
W''(\tilde{\sigma};z^X) = \left. \frac{d^2 W}{d\sigma^2}(\sigma;z^X) \right|_{\tilde{\sigma}}
\uqe
Following our earlier observations for the FJ Hamiltonians let us focus on the prediction for the correlation exactly at the position of the spin-plane for the model in question. In this case (\ref{csig}) reduces to
\equ
G_0^{(\sigma)}(z^X,z^X;z^X) = \frac{1}{W''(\tilde{\sigma};z^X)}
\uqe
which can be explicitly calculated using techniques very similar to those described in Sec.\ 2.2 of \cite{pb}. This procedure identically reproduces the analytic expression for the MF correlation function so that we identify
\equ
G^{(\sigma)}_0(z^X,z^X;z^X) = G_0(z^X,z^X)
\uqe
In this way we can reconstruct the MF correlation function from a set of spin-like Hamiltonians $\{ H[\sigma;z^X] \}$ using each element at the appropriate local position in space. Similar remarks can be shown to be valid to order $q^2$ (and beyond if rigidity-like terms are allowed for in (\ref{HS})) so we may identify
\equ
G(z^X,z^X;{\bf q}) = \frac{1}{W''(\tilde{\sigma};z^X) + \Sigma(\tilde{\sigma};z^X)q^2 + \cdots }
\uqe
and recall that $\tilde{\sigma} = \tilde{m}(z^X)$.        

We can easily generalise the analysis to derive effective Hamiltonians containing any number of spin- and interfacial-like collective coordinates by imposing spin and crossing constraints at arbitrary positions. Of particular interest is the two-field model $H[\sigma,\ell]$ with $\sigma({\bf y})$ the $d-1$ dimensional plane of spins at the wall and $\ell({\bf y})$ the collective coordinate for the surface of fixed magnetization $m^X=0$. This is a coupled Hamiltonian similar to the $H[\ell_1,\ell_2]$ theory describing the interaction between order parameter fluctuations near the interface and wall. The model can be constructed from the properties of the planar constrained profile $m_\pi(z;\sigma_\pi,\ell_\pi)$ which satisfies the usual Euler-Lagrange equation subject to the constraints
\eqan{sigconX}
m_\pi(0;\sigma_\pi,\ell_\pi) &=& \sigma_\pi \\
m_\pi(\ell_\pi;\sigma_\pi,\ell_\pi) &=& 0
\naqe
using now standard methods. We simply quote the results since the details of a more involved calculation will be discussed later. The Hamiltonian is
\eqan{Hsigma}
H[\sigma,\ell] &=& \int d{\bf y} \biggl\{ \frac{1}{2} \Sigma_{11} (\nabla \sigma)^2 + \Sigma_{12} \nabla \sigma . \nabla \ell \nonu \\
& & + \frac{1}{2} \Sigma_{22} (\nabla \ell)^2 + W_2(\sigma,\ell) \biggr\}
\naqe
where explicit evaluation of the various terms in the reliable double parabola approximation for $\phi(m)$ yields
\eqan{wsig1}
W_2(\sigma,\ell) &=& \frac{1}{2} r (\sigma-\sigma_0)^2 +  2\kappa m_\alpha (\sigma-m_\alpha) {\rm e}^{-\kappa \ell} \nonu \\
& & + \kappa \left( m^2_\alpha +(\sigma-m_\alpha)^2 \right) {\rm e}^{-2\kappa \ell}\naqe
with  
\eqan{wsig2}
r &=& c+\kappa \nonu \\
\sigma_0 & = & m_\alpha + \tau
\naqe                        
                    
Similarly the stiffness coefficients are given by 
\eqa
\Sigma_{11} & = & \left. \int_0^\infty dz \left( \frac{\partial m_\pi}{\partial \sigma_\pi} \right)^2 \right|_{\sigma({\bf y}),\ell({\bf y})} \nonu \\
 & = & \frac{1}{2 \kappa} + \frac{{\rm e}^{-2\kappa \ell}}{\kappa} \\ 
\Sigma_{12} & = & \left. \int_0^\infty dz \left( \frac{\partial m_\pi}{\partial \sigma_\pi} \right) \left( \frac{\partial m_\pi}{\partial \ell_\pi} \right)
\right|_{\sigma({\bf y}),\ell({\bf y})}  \nonu \\
& = & m_\alpha(\kappa \ell - 1) {\rm e}^{-\kappa \ell} +O({\rm e}^{-\kappa \ell})  
\aqe
and
\eqa                                                                           \Sigma_{22} & = &  \left. \int_0^\infty dz \left( \frac{\partial m_\pi}{\partial \ell_\pi} \right)^2 \right|_{\sigma({\bf y}),\ell({\bf y})} \nonu \\ 
&=& \Sigma_{\alpha \beta} +2\kappa m_\alpha(\sigma-m_\alpha) {\rm e}^{-\kappa \ell} - 2 \kappa^2 m_\alpha^2 \ell {\rm e}^{-2\kappa \ell} \nonu \\
& & + \kappa \left[ 2(\sigma-m_\alpha)^2 + 3 m_\alpha^2 \right] {\rm e}^{-2\kappa \ell} 
\aqe                                                                           Note that minimization of $W(\sigma,\ell)$ recovers the MF surface magnetization and position of the $\alpha \beta$ interface. The two-field model $H[\sigma,\ell]$ is an alternative candidate for describing coupling effects at wetting transitions and has its own CFRS which enable us to address problem (P3). We again simply quote the results since a more general proof will be given in the next section.         

First construct the appropriate direct correlation function matrix
\equ
{\bf C}({\bf q}) = \left( \begin{array}{ll} \partial^2_{11} & \partial^2_{12} \\ \partial^2_{21} & \partial^2_{22} \end{array} \right) W_2(\sigma,\ell) + q^2 {\bf \Sigma}
\uqe
where this time
\equ
\begin{array}{lclcl} \partial^2_{11} = \frac{\partial^2}{\partial \sigma^2} &;& \partial^2_{12}=\partial^2_{21}= \frac{\partial^2}{\partial \sigma \partial \ell} &;& \partial^2_{22} = \frac{\partial^2}{\partial \ell^2} \end{array}
\uqe
and all derivatives are evaluated at equilibrium. The required MF order parameter correlation functions are then given by
\eqan{cfrssig}
G(0,0;{\bf q}) &=& S_{11}({\bf q}) \nonu \\
G(0,\tilde{\ell};{\bf q}) &=& -\tilde{m}'(\tilde{\ell}) S_{12}({\bf q}) \nonu \\
G(\tilde{\ell},\tilde{\ell};{\bf q}) &=& \tilde{m}'(\tilde{\ell})^2 S_{22}({\bf q})
\naqe                                                                           where the structure factor matrix ${\bf S}({\bf q})$ (defined analogous to (\ref{sdef})) is the inverse of ${\bf C}({\bf q})$ and $\tilde{m}(\tilde{\ell})=0$.   

Let us finish this subsection by using the $H[\sigma,\ell]$ Hamiltonian to calculate the non-Lorentzian form of $G(0,0;{\bf q})$ for critical wetting along route (i). For simplicity we shall ignore the position dependence of the stiffness coefficients although these can easily be included (see Sec.\ \ref{critwetref} for further discussion).    

The elements of the matrix ${\bf C}$ can be calculated from (\ref{wsig1}) and (\ref{wsig2}). To the appropriate order we find (where $\nu = \frac{c-\kappa}{c+\kappa}$)
\eqa
C_{11}(0) &=& c+\kappa + \frac{2 \kappa \nu^2 \tau^2}{m_\alpha^2} \\
C_{12}(0) &=& 2 \kappa^2 \nu \tau \\
C_{22}(0) &=& 2 \kappa^3 \nu^2 \tau^2
\aqe
and so using (\ref{cfrssig}) derive
\equ
G(0,0;{\bf q}) = \frac{1}{c+\kappa + \frac{q^2}{2\kappa} - \frac{2\kappa}{1+q^2 \xi_\parallel^2}}
\uqe
for $\xi_\parallel^2 = \frac{m^2_\alpha \nu}{2\kappa^2 \tau^2}$. This is of the desired form consistent with (\ref{Gcw}) and the exact sum-rules (\ref{sumrules}) quoted earlier. 

In fact the form of $G(0,0;{\bf q})$ at critical wetting emerges much more naturally from the CFRS of $H[\sigma,\ell]$ model compared to that of the PB model because in the latter we have to explicitly include the vanishing of the magnetization gradient at the wall (recall (\ref{gsr})). In the $H[\sigma,\ell]$ description the singularity of $\tilde{m}'_1$ is implicitly included in the binding potential $W_2(\sigma,\ell)$. This feature works against us somewhat if we use the $H[\sigma,\ell]$ model to calculate $G(0,0;{\bf q})$ at complete wetting. While we can derive a formally correct expression for this quantity, comparison with (\ref{PBresult}) is not as immediate as with the $H[\ell_1,\ell_2]$ approach which has the natural inclusion of explicit  $\tilde{m}_1^{'2}$ terms in its CFRS. Consequently there is no elegant analogue of the stiffness matrix free energy relation in the $H[\sigma,\ell]$ formalism of complete wetting. Thus, while both Hamiltonians $H[\sigma,\ell]$ and $H[\ell_1,\ell_2]$ yield precisely equivalent expressions for the MF correlation function, which formally solves the Ornstein-Zernike equation (\ref{ozdif}), the spin- and interfacial-like representations of order parameter fluctuations at the wall are more convenient for critical and complete wetting respectively. We shall return to this point later.

\subsection{Invariance of correlations within the CFRS}
In the section above we stated that the two different collective coordinate models $H[\sigma,\ell]$ and $H[\ell_1,\ell_2]$ constitute equivalent approaches to recovering MF correlation functions (near the wall and $\alpha \beta$ interface). That is, apart from the question of convenience for analysing particular transitions both choices of collective coordinate give the same final expression for the physical order parameter correlation functions $G(z_\mu,z_\nu;{\bf q})$ at MF level. Here we give a general proof of the invariance of the $G(z_\mu,z_\nu;{\bf q})$ with respect to different effective Hamiltonian descriptions. This section is rather technical and the reader may prefer to skip to Sec.\ \ref{summary} where we summarise the main points.     

Let us denote the collective coordinates $X_1({\bf y})$ and $X_2({\bf y})$ which we use to represent order parameter fluctuations at the wall and interface respectively. The proof can be trivially extended to any number of fields at arbitrary positions. Associated with the collective coordinates is a pair of surfaces described by $Z_1({\bf y})$ and $Z_2({\bf y})$. These may be viewed as the locations of the fluctuating fields and are specified by appropriate functions
\equ
Z_i({\bf y}) = Z_i \Bigl( X_1({\bf y}), X_2({\bf y}) \Bigr)
\uqe
which will depend on the choice of collective coordinate system.      

The magnetization profile is doubly constrained by the conditions
\equn{con2}
m \Bigl( {\bf y},Z_i({\bf y});X_1,X_2 \Bigr) = M_i \Bigl( X_1({\bf y}), X_2({\bf y}) \Bigr) 
\nuqe
and notice that we have allowed the value of the magnetization at position $Z_i({\bf y})$ to depend on one or indeed both of the collective coordinates. The effective Hamiltonian is defined in the usual way as a constrained fluctuation sum  
\equ
H[X_1,X_2] = - \ln \left( \int_C {\cal D} m {\rm e}^{-H_{LGW}[m]} \right)
\uqe
We will assume that it is possible to find collective coordinates $X_1, X_2$ such that the FJ saddle point approximation is valid. Then
\equn{gHdefn2}
H[X_1,X_2] = H_{LGW}[m_\Xi] 
\nuqe
where $m_\Xi$ is the profile that minimises $H_{LGW}$ subject to (\ref{con2}). Thus $m_\Xi$ satisfies the usual Euler-Lagrange equation (\ref{2el}) together with the required boundary conditions (\ref{2bc}). Following all our other treatments we calculate $m_\Xi$ perturbatively in terms of the planar constrained profile which satisfies the analogue of (\ref{con2})
\equn{cons2planar}
m_\pi(Z_{i\pi};X_{1\pi},X_{2\pi}) = M_i(X_{1\pi},X_{2\pi})
\nuqe
with $Z_{i \pi} = Z_i( X_{1\pi},X_{2\pi})$. Following Fisher, Jin and Parry \cite{fjp} we write
\eqan{ansz}
m_\Xi \Bigl( {\bf y},z;X_1,X_2 \Bigr) &=& m_\pi \Bigl( z;X_1({\bf y}),X_2({\bf y}) \Bigr) \nonu \\
& & + {\cal E} \Bigl( {\bf r}; X_1,X_2 \Bigr)
\naqe
where ${\cal E}({\bf r};X_1,X_2)$ denotes the error term. By construction ${\cal E}$ vanishes for planar configurations       
\equ
X_i({\bf y}) = X_{i\pi} \mbox{\ \ \ \ \ $\forall {\bf y}$} 
\uqe
and also at the position of the collective coordinates 
\equ
z=Z_i({\bf y})
\uqe
for any ${\bf y}$.      

Now let us denote the equilibrium MF values of the fields $\tilde{X}_1$ and $\tilde{X}_2$ respectively. These may be found by minimisation of the Hamiltonian $H[X_1,X_2]$. Clearly setting $X_i=\tilde{X}_i$ recovers the MF profile  
\equn{conscon}
m_\pi(z;\tilde{X}_1,\tilde{X}_2) = \tilde{m}(z) 
\nuqe
and together with (\ref{con2}) this implies 
\equ
\tilde{m}(\tilde{Z}_i) = M_i(\tilde{X}_1,\tilde{X}_2)
\uqe
with $\tilde{Z}_i$ the MF position of the collective coordinate $X_i$.   

To elucidate the reconstruction scheme we need to find a relationship between the magnetization pair correlation functions of the LGW theory and the collective coordinate correlations of the effective Hamiltonian. Recall that the former are defined by
\eqa
G({\bf r}_1,{\bf r}_2) &=& G(z_1,z_2;{\bf y}_2-{\bf y}_1) \nonu \\
& =& \langle m({\bf r}_1) m({\bf r}_2) \rangle - \langle m({\bf r}_1)\rangle \langle m({\bf r}_2) \rangle
\aqe 
and satisfy the Ornstein-Zernike equation \cite{amit}
\equn{ozeqn}
\int d{\bf r} \left. \frac{ \delta^2 H_{LGW}[m]}{\delta m({\bf r}_1) \delta m({\bf r})} \right|_{m=\tilde{m}} G({\bf r},{\bf r}_2) = \delta({\bf r}_1 - {\bf r}_2) 
\nuqe
 Similarly, for the macroscopic collective coordinate theory
\eqan{Sagain}
S({\bf y}_1,{\bf y}_2) &=& S({\bf y}_2 -{\bf y}_1) \nonu \\
&=& \langle \ell({\bf y}_1) \ell({\bf y}_2) \rangle - \langle \ell({\bf y}_1) \rangle \langle \ell({\bf y}_2) \rangle
\naqe
which obey \cite{amit}
\eqan{ozeqn2}
& \int d{\bf y} \left. \frac{\delta^2 H[X_1,X_2]}{\delta X_\mu({\bf y}_1) \delta X_\rho({\bf y})} \right|_{X_1=\tilde{X_1},X_2=\tilde{X_2}} S_{\rho \nu}({\bf y},{\bf y}_2) = & \nonu \\
& \delta_{\mu \nu} \; \delta({\bf y}_1 - {\bf y}_2) &
\naqe                                                  
and note our implicit use of the summation convention.      

Considering the Ornstein-Zernike direct correlation function
\equ
\frac{ \delta^2 H_{LGW}[m_\Xi]}{\delta m_\Xi({\bf r}) \delta m_\Xi({\bf r}')} 
\uqe
with an eye on (\ref{gHdefn2}), the correlation functions $S({\bf y}_1,{\bf y}_2)$ can be introduced by changing the functional derivatives to be with respect to $X_i$ rather than $m_\Xi$. For this we need to use the functional chain rule
\eqa
\lefteqn{\frac{\delta}{\delta m_\Xi({\bf y},z)} =} \nonu \\
& &  \int d{\bf y}_1 \Biggl\{ \frac{\delta X_1({\bf y}_1)}{\delta m_\Xi({\bf r})} \frac{\delta}{\delta X_1({\bf y}_1)} + \frac{\delta X_2({\bf y}_1)}{\delta m_\Xi({\bf r})} \frac{\delta}{\delta X_2({\bf y}_1)} \Biggr\}
\aqe
so that we can write
\eqan{above}
\lefteqn{\frac{ \delta^2 H_{LGW}[m_\Xi]}{\delta m_\Xi({\bf r}) \delta m_\Xi({\bf r}')} =} \nonu \\
& & \int d{\bf y}_1 \Biggl\{ \frac{\delta X_1({\bf y}_1)}{\delta m_\Xi({\bf r})} \frac{\delta}{\delta X_1({\bf y}_1)} + \frac{\delta X_2({\bf y}_1)}{\delta m_\Xi({\bf r})} \frac{\delta}{\delta X_2({\bf y}_1)} \Biggr\} \nonu \\
& & \times \int d{\bf y}_2 \Biggl\{ \frac{\delta X_1({\bf y}_2)}{\delta m_\Xi({\bf r}')} \frac{\delta}{\delta X_1({\bf y}_2)} + \frac{\delta X_2({\bf y}_2)}{\delta m_\Xi({\bf r}')} \frac{\delta}{\delta X_2({\bf y}_2)} \Biggr\} \nonu \\
& & \times H[X_1,X_2]  
\naqe                                                     
Fortunately this can be written in matrix form. Defining
\eqa
x({\bf y},{\bf r}) & = & \left[ \begin{array}{c} \frac{\delta X_1({\bf y})}{\delta m_\Xi({\bf r})} \\ \frac{\delta X_2({\bf y})}{\delta m_\Xi({\bf r})} \end{array} \right] \\
C({\bf y}_1,{\bf y}_2) & = & \left[ \begin{array}{ll} 
\frac{\delta^2 H[X_1,X_2]}{\delta X_1({\bf y}_1) \delta X_1({\bf y}_2)} & \frac{\delta^2 H[X_1,X_2]}{\delta X_1({\bf y}_1) \delta X_2({\bf y}_2)} \\
\frac{\delta^2 H[X_1,X_2]}{\delta X_2({\bf y}_1) \delta X_1({\bf y}_2)} & \frac{\delta^2 H[X_1,X_2]}{\delta X_2({\bf y}_1) \delta X_2({\bf y}_2)} \end{array} \right] 
\aqe
we can rewrite (\ref{above})
\eqan{Cmatrix}
\lefteqn{\left. \frac{\delta^2 H_{LGW}[m]}{\delta m({\bf r}) \delta m({\bf r'})} \right|_{m=m_\Xi} =} \nonu \\
& & \int d{\bf y}_1 d{\bf y}_2 x^T({\bf y}_1,{\bf r}) C({\bf y}_1,{\bf y}_2) x({\bf y}_2,{\bf r'}) 
\naqe
The elements of the effective Hamiltonian direct correlation function matrix ${\bf C}({\bf y}_1,{\bf y}_2)$ are                                               \equn{Cele}
C_{\mu \nu}({\bf y}_1,{\bf y}_2) \equiv \frac{\delta^2 H[X_1,X_2]}{\delta X_\mu({\bf y}_1) \delta X_\nu({\bf y}_2)} = C_{\mu \nu}({\bf y}_1-{\bf y}_2)
\nuqe     
with derivatives are evaluated at equilibrium $X_i=\tilde{X}_i$. 

So far we have succeeded in writing the first part of the Ornstein-Zernike equation (\ref{ozeqn}) in matrix form involving functions of the effective theory. It is a simple matter to show that the magnetization correlations can be expressed in a similar way. Define
\equ
w({\bf y},{\bf r})  =  \left[ \begin{array}{c} \frac{\delta m_\Xi({\bf r})}{\delta X_1({\bf y})} \\ \frac{\delta m_\Xi({\bf r})}{\delta X_2({\bf y})} \end{array} \right]
\uqe
then
\equn{Gmatrix}
G({\bf r},{\bf r'}) = \int d{\bf y}_1 d{\bf y}_2 w^T({\bf y}_1,{\bf r}) S({\bf y}_1,{\bf y}_2) w({\bf y}_2,{\bf r'}) 
\nuqe
where $S$ is the functional inverse of $C$
\equn{Sdefn}
\int d{\bf y} C_{\mu \rho} ({\bf y}_1,{\bf y}) S_{\rho \nu}({\bf y},{\bf y}_2) = \delta_{\mu \nu} \delta({\bf y}_1 - {\bf y}_2) 
\nuqe    
Clearly $S$ is the matrix of correlation functions $S_{\mu \nu}({\bf y}_1,{\bf y}_2)$. To prove (\ref{Gmatrix}) notice that the Ornstein-Zernike equation can be written
\eqan{ozagain}
\delta({\bf r}_1-{\bf r}_2) & = & \int d{\bf r} d{\bf u}_1 d{\bf u}_2 d{\bf v}_1 d{\bf v}_2 \;  x^T({\bf u}_1, {\bf r}_1) C({\bf u}_1,{\bf u}_2) \nonu \\
& & \times x({\bf u}_2,{\bf r}) w^T({\bf v}_1,{\bf r}) S({\bf v}_1,{\bf v}_2) w({\bf v}_2,{\bf r}_2)
\naqe                                                  
where we have used (\ref{Cmatrix}) and (\ref{Gmatrix}). Next from the chain rule identity
\equ
 \int \frac{\delta X_{\mu}({\bf y})}{\delta m_\Xi({\bf r})} \frac{\delta m_\Xi({\bf r})}{\delta X_{\nu}({\bf x})} d{\bf r} = \frac{\delta X_\mu({\bf y})}{\delta X_{\nu}({\bf x})} =  \delta({\bf y}-{\bf x}) \delta_{\mu \nu}
\uqe
one observes that
\eqa
\int d{\bf r} x({\bf u}_2,{\bf r}) w^T({\bf v}_1,{\bf r}) & = & \int \left[ \begin{array}{c} \frac{\delta X_1({\bf u}_2)}{\delta m_\Xi({\bf r})} \\ \frac{\delta X_2({\bf u}_2)}{\delta m_\Xi({\bf r})} \end{array} \right] \nonu \\
& & \times \left[ \begin{array}{ll} \frac{\delta m_\Xi({\bf r})}{\delta X_1({\bf v}_1)} & \frac{\delta m_\Xi({\bf r})}{\delta X_2({\bf v}_1)} \end{array}  \right] d{\bf r} \nonu \\
& = & \left[ \begin{array}{ll} 
1 & 0 \\
0 & 1 \end{array} \right] \delta({\bf u}_2-{\bf v}_1) 
\aqe
Then, from (\ref{Sdefn}) the r.h.s.\ of (\ref{ozagain}) becomes
\eqan{back}
\int d{\bf u}_1 x^T({\bf u}_1,{\bf r}_1) w({\bf u}_1,{\bf r}_2) & = & \int d{\bf u}_1 \left[ \begin{array}{ll} \frac{\delta X_1({\bf u}_1)}{\delta m_\Xi({\bf r}_1)} & \frac{\delta X_2({\bf u}_1)}{\delta m_\Xi({\bf r}_1)} \end{array}  \right] \nonu \\
& & \times \left[ \begin{array}{c} \frac{\delta m_\Xi({\bf r}_2)}{\delta X_1({\bf u}_1)} \\ \frac{\delta m_\Xi({\bf r}_2)}{\delta X_2({\bf u}_1)} \end{array}  \right] \nonumber \\
& = & \delta({\bf r}_1 - {\bf r}_2) 
\naqe
using the identity
\eqa
& \int d{\bf y} \left\{ \frac{ \delta X_1({\bf y}) }{ \delta m_\Xi({\bf r}_1) } \frac{ \delta m_\Xi({\bf r}_2) }{ \delta X_1({\bf y}) } + \frac{ \delta X_2({\bf y}) }{ \delta m_\Xi({\bf r}_1) } \frac{ \delta m_\Xi({\bf r}_2) }{ \delta X_2({\bf y}) } \right\} = &\nonu \\
& \delta({\bf r}_1-{\bf r}_2) & 
\aqe
                                                      
Equation (\ref{back}) is just the l.h.s.\ of (\ref{ozagain}) and consequently the Ornstein-Zernike relation is satisfied. Therefore we have shown that 
\equn{ozeqnag}
\left. \int d{\bf r} \frac{ \delta^2 H_{LGW}[m]}{\delta m({\bf r}_1) \delta m({\bf r})} \right|_{m=m_\Xi} G({\bf r},{\bf r}_2) = \delta({\bf r}_1 - {\bf r}_2)
\nuqe                                                    
If we set $X_i({\bf y})= \tilde{X}_i$ the consistency condition (\ref{conscon}) together with (\ref{ansz}) imply that the above relation reduces to the required (\ref{ozeqn}). In this way the pair correlation function (in MF approximation) is
\equn{Gatlast}
G({\bf r},{\bf r'}) = \left. \int d{\bf y}_1 d{\bf y}_2 w^T({\bf y}_1,{\bf r}) S({\bf y}_1,{\bf y}_2) w({\bf y}_2,{\bf r'}) \right|_{X_i=\tilde{X}_i} 
\nuqe
                                                      
To simplify (\ref{Gatlast}) we first functionally differentiate (\ref{ansz}) with respect to $X_\mu$
\equ
\frac{\delta m_\Xi({\bf r})}{\delta X_\mu({\bf y}')} = \delta({\bf y}-{\bf y}') \frac{\partial m_\pi}{\partial X_{\mu \pi}} (z; X_1({\bf y}),X_2({\bf y})) + \frac{\delta {\cal E} ({\bf r})}{\delta X_\mu({\bf y}')} 
\uqe
This implies                                                   
\eqan{upabove}
G({\bf r},{\bf r'})& =& \int d{\bf y}_1 d{\bf y}_2 \left[ \begin{array}{ll} \frac{\partial m_\pi}{\partial X_{1\pi}}(z) & \frac{\partial m_\pi}{\partial X_{2\pi}}(z) \end{array}  \right]  \delta({\bf y}-{\bf y}_1) 
\nonu \\
& & \times S({\bf y}_1,{\bf y}_2)  \left[ \begin{array}{c} \frac{\partial m_\pi}{\partial X_{1\pi}}(z') \\ \frac{\partial m_\pi}{\partial X_{2\pi}}(z') \end{array}  \right] \delta({\bf y}'-{\bf y}_2) \nonu \\
& & + \left.  O \left(\frac{\delta {\cal E}}{\delta X_\mu} \right) \right|_{\tilde{X}_1,\tilde{X}_2} 
\naqe
Here the r.h.s.\ is evaluated at the planar values  $X_1 = \tilde{X}_1$ and $X_2=\tilde{X}_2$ implying that the error term  ${\cal E}$ vanishes. However it is not necessary that the $\frac{\delta {\cal E}}{\delta X_\mu}$ also vanishes. In fact it only does so at particular points as we now show. 

As remarked earlier the Fisher-Jin-Parry ansatz (\ref{ansz}) is constructed so that the error term vanishes when $z=Z_i({\bf y})$ i.e.
\equ
{\cal E} \Bigl( {\bf y},Z_i({\bf y});X_1,X_2 \Bigr) = 0 
\uqe 
Recalling that $Z_i({\bf y}) = Z_i(X_1({\bf y}),X_2({\bf y}))$ functional differentiation with respect to $X_\mu({\bf x})$ gives                                                       
\equ
\left. \frac{\partial {\cal E}}{\partial z}({\bf y},z) \right|_{z=Z_i({\bf y})} \frac{\delta Z_i({\bf y})}{\delta X_\mu({\bf x})} + \left. \frac{\delta {\cal E}({\bf y},z)}{\delta X_\mu({\bf x})} \right|_{z=Z_i({\bf y})} =0
\uqe
implying
\equn{err}
\left. \frac{\delta {\cal E}({\bf y},z)}{\delta X_\mu({\bf x})} \right|_{z=Z_i({\bf y})} = - \delta({\bf x}-{\bf y}) \frac{\partial Z_i}{\partial X_\mu}({\bf y}) . \left. \frac{\partial {\cal E}}{\partial z}({\bf y},z) \right|_{z=Z_i({\bf y})} 
\nuqe
Evaluated at the MF values of the fields $\tilde{X}_i$ the r.h.s.\ disappears since ${\cal E}$ vanishes identically and hence so do its partial derivatives, implying
\equ
\left. \frac{\delta {\cal E}({\bf y},z)}{\delta X_\mu({\bf x})} \right|_{z=\tilde{Z}_i, X_i=\tilde{X}_i} \propto \left. \frac{\partial {\cal E}}{\partial z} \right|_{z=\tilde{Z}_i, X_i=\tilde{X}_i} = 0 
\uqe
Therefore the error term in (\ref{upabove}) only vanishes for 
\equ
z,z' \in \{ \tilde{Z}_1, \tilde{Z}_2 \} 
\uqe
Multiplying out the matrices in (\ref{upabove}) and choosing $z$ and $z'$ according to the above criterion one finally arrives at
\eqan{CFRS}
\lefteqn{ G(\tilde{Z}_i,\tilde{Z}_j;{\bf y}-{\bf y}') =} \nonu \\
& &  \left. \frac{\partial m_\pi}{\partial X_{\mu \pi}} (\tilde{Z}_i) \frac{\partial m_\pi}{\partial X_{\nu \pi}} (\tilde{Z}_j) S_{\mu \nu}({\bf y}-{\bf y}') \right|_{\tilde{X}_1,\tilde{X}_2} 
\naqe
which is the central result of the generalized CFRS. It follows that for any two collective coordinates satisfying the local constraints (\ref{con2}) it is possible to recover the underlying MF correlations of the LGW theory for particle positions at and between the planes $\tilde{Z}_i$ (corresponding to the MF locations of the fields). Connection with our earlier statements can be made by simply taking the Fourier transform with respect to the transverse variable. The relation establishes that the physical order parameter correlation function is an invariant, independent of the choice of collective coordinates.     

Before we use the formalism to construct effective Hamiltonians based on collective coordinates with both spin- and interfacial-like components we illustrate how the general CFRS reduces to the matrix equations for the models considered earlier.

\subsubsection*{Examples} 

\setcounter{c}{0}
\begin{list}{(\roman{c})}{\usecounter{c} }
\item The PB model         

For this case the collective coordinates $X_i({\bf y}) = \ell_i({\bf y})$ are interfacial-like and may be simply identified with positions $Z_i({\bf y})$    
\equ
X_i({\bf y}) = \ell_i({\bf y}) = Z_i({\bf y})
\uqe                                                       
and the planar constraint is
\equn{pbcons}
m_\pi(\ell_{i\pi};\ell_{1\pi},\ell_{2\pi}) = m^X_i 
\nuqe
This leads to a simplification of the $\frac{\partial m_\pi}{\partial X}$ terms in (\ref{CFRS}). Differentiation yields
\equ
\frac{\partial m_\pi}{\partial \ell_{j\pi}}(z;\ell_{1\pi},\ell_{2\pi}) = 0 \mbox{\hspace*{5mm} for $z=\ell_{i\pi}$} 
\uqe
and
\eqa
\frac{\partial m_\pi}{\partial \ell_{i\pi}}(z;\ell_{1\pi},\ell_{2\pi}) &=& - \left. \frac{\partial m_\pi}{\partial z}(z;\ell_{1\pi},\ell_{2\pi}) \right|_{z=\ell_{i\pi}} \nonu \\
& = & -\tilde{m}'(\tilde{\ell}_i) \mbox{\hspace*{5mm} for $\ell_{i\pi}=\tilde{\ell}_i$} 
\aqe
and recall that the equilibrium positions $\tilde{\ell}_i \equiv z_i$ are solutions of $m^X_i=\tilde{m}(z)$. Substitution into the invariance relation (\ref{CFRS}) and taking a Fourier transform yields
\eqan{2field}
G(z_1,z_1;{\bf q}) & = & \tilde{m}'(z_1)^2 S_{11}({\bf q}) \nonumber \\
G(z_2,z_2;{\bf q}) & = & \tilde{m}'(z_2)^2 S_{22}({\bf q}) \nonumber \\
G(z_1,z_2;{\bf q}) & = & \tilde{m}'(z_1)\tilde{m}'(z_2) S_{12}({\bf q})
\naqe
as quoted earlier in Sec.\ \ref{twofield}.

\item The spin-like Hamiltonian        

The collective coordinates are $X_1=\sigma$, $X_2=\ell$ and correspond to positions $Z_1=0$, $Z_2=\ell$ respectively. From the constraint equations (\ref{sigconX}) we can calculate the  $\frac{\partial m_\pi}{\partial X}$ terms which satisfy
\equ
\frac{\partial m_\pi}{\partial \sigma_\pi}(\ell_\pi;\sigma_\pi,\ell_\pi) = \frac{\partial m_\pi}{\partial \ell_\pi}(0;\sigma_\pi,\ell_\pi) = 0
\uqe 
with                                                       
\eqa
\frac{\partial m_\pi}{\partial \ell_\pi}(z;\sigma_\pi,\ell_\pi) &=& \left.  -\frac{\partial m_\pi}{\partial z}(z;\sigma_\pi,\ell_\pi) \right|_{z=\ell_\pi} \nonu \\
&=& -\tilde{m}'(\tilde{\ell})  
\aqe
at equilibrium ($\ell_{\pi}=\tilde{\ell}$ and $\sigma_\pi=\tilde{\sigma}$). The CFRS equations (\ref{cfrssig}) then follow directly from the general result (\ref{CFRS}) if we also use (\ref{sigdif}).
\end{list}

\subsection{Summary}   
\label{summary}
The general CFRS provides the necessary mathematical framework for a complete solution to problem (P3). For arbitrary collective coordinates $X_1$, $X_2$ with MF positions $z_1$ (near the wall) and $z_2$ (near the $\alpha \beta$ interface) we can construct an effective Hamiltonian using a saddle point identification
\equ
H[X_1,X_2] = \min_C H_{LGW}[m]
\uqe
subject to appropriate constraints. The MF order parameter correlation functions (satisfying the Ornstein-Zernike equation (\ref{ozdif})) can be recovered using the invariant relation (noting the summation convention)
\equ
G(z_\mu,z_\nu;{\bf q}) = \frac{\partial m_\pi}{\partial X_{\mu'}} (z_\mu;\{ X_i \}) \frac{\partial m_\pi}{\partial X_{\nu'}} (z_\nu; \{ X_i \} ) S_{\mu' \nu'}({\bf q}) 
\uqe
where $m_\pi(z; \{ X_i \})$ are the planar constrained profiles. The matrix ${\bf S}$ has elements                                                        
\equ
S_{\mu \nu}({\bf q}) = \int d{\bf y} {\rm e}^{i {\bf q}.{\bf y}} \langle \delta X_\mu({\bf y}) \delta X_\nu({\bf 0}) \rangle
\uqe
corresponding to generalized structure factors. In practise we will only be interested in Hamiltonians modelling long wavelength fluctuations in which case the models all have the form
\eqa
H[X_1,X_2] &=& \int d{\bf y} \biggl\{ \frac{1}{2} \Sigma_{\mu \nu}(X_1,X_2) \nabla X_\mu . \nabla X_\nu \nonu \\
& &  + W_2(X_1,X_2) \biggr\}
\aqe
with generalized stiffness coefficients and binding potential determined by $m_\pi(z; \{ X_i \})$. The formulae for these are given by obvious generalisations of (\ref{GEN1}) and (\ref{GEN2}) replacing the $\ell_i$ by $X_i$. The structure factor matrix ${\bf S}$ is the inverse of the direct matrix 
\equ
{\bf C}({\bf q}) = \int d{\bf y}_{12} {\rm e}^{i {\bf q}.{\bf y}_{12}} {\bf C}({\bf y}_{12})
\uqe
with
\equ
C_{\mu \nu}({\bf y}_{12}) = \frac{\delta^2 H[X_1,X_2]}{\delta X_\mu({\bf y}_1) \delta X_\nu({\bf y}_2)} 
\uqe

Focusing on long wavelength fluctuations the matrix ${\bf C}({\bf q})$ has elements
\equ
C_{\mu \nu}({\bf q}) = \frac{\partial^2 W(X_1,X_2)}{\partial X_\mu \partial X_\nu} + q^2 \Sigma_{\mu \nu} (X_1,X_2)
\uqe 
ignoring rigidity terms of $O(q^4)$ which can also be calculated if so wished.      
This scheme is valid for general collective coordinates $(X_1, X_2)$ such as the PB choice $(\ell_1, \ell_2)$ and spin-like model $(\sigma,\ell)$ discussed at length earlier. If our sole problem was (P3), that is to develop a effective Hamiltonian theory of correlations at the wall, the only remaining concern would be which choice of coordinates is most convenient for studying critical and complete wetting, say. In fact we have already answered this. At a complete wetting transition occurring for temperatures $T\gg T_W$ the CFRS is best studied using the PB model $H[\ell_1,\ell_2]$ because the surface gradient $\tilde{m}'_1$ is large and does not exhibit any singular as $h \goto 0^-$. On the other hand for critical wetting it is more convenient to use the $H[\sigma,\ell]$ model because $\tilde{m}_1'$ contains important scaling behaviour and vanishes at $T=T_W$ for $h=0$. It is therefore better to adopt a CFRS in which the $\tilde{m}'_1$ do not enter explicitly but are implicitly contained in the binding potential.

These remarks are strongly suggestive that we introduce a model based on a collective coordinate $X_1$ describing order parameter fluctuations near the wall that can smoothly interpolate between spin-like ($\sigma$) and interfacial-like ($\ell$) representations. With such a continuous family of possible collective coordinates we will be better placed to ask which choice is optimal for describing coupling effects beyond MF approximation.

\section{Optimal two-field theory}
\label{optimal}

\subsection{Proper collective coordinates}   
\begin{figure}[h]
\begin{center}
\scalebox{0.55}{\includegraphics{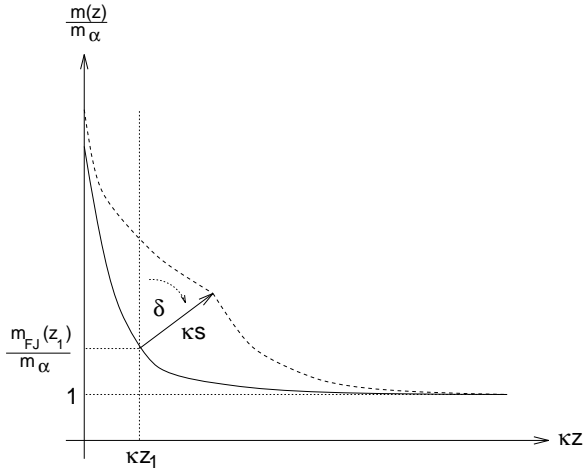}}
\end{center}
\caption{Detail of the planar magnetization profiles near the wall in scaled units. The broken curve shows $m(z)$  which incorporates a local enhancement and translation of the FJ profile (corresponding to the solid line). The proper coordinate $s$ and angle $\delta$ are shown.}
\label{euro}
\end{figure}

First let us state that the choice of upper field $X_2$ is not in question and that fluctuations in the local position of the $\alpha \beta$ interface will be described by the interfacial-like collective coordinate $\ell({\bf y})$ corresponding to the surface of fixed magnetization $m^X=0$. For a given value of (planar) field $\ell_\pi$ let us denote the FJ profile
\equ
m_\pi(z;\ell_\pi) \equiv m^{FJ}(z;\ell_\pi)
\uqe
which will simplify our notation. Now the FJ theory does not account for fluctuations near the wall and it is natural to introduce a collective coordinate $X_1$ which perturbs their profile in this area. We will denote the collective coordinate $s$ and allow it to have both interfacial- and spin-like components (note that while the upper field certainly has a spin component nothing is lost by just including the much much larger interfacial fluctuations). This may be achieved by supposing that the doubly constrained profile is locally translated {\it and} enhanced. Specifically, choose an arbitrary position $z_1$ close to the wall satisfying $0 \le z_1 \lesssim \kappa^{-1}$. The value of $z_1$ will not appear in our final equations concerning problems (P1) and (P2) and in calculations we can set it equal to zero without loss of any information. Next we construct a doubly constrained (planar) profile which satisfies
\equn{GCon}
m_\pi(z_1+\ell_{1\pi};s_\pi,\ell_\pi) = m^{FJ}(z_1;\ell_\pi) + \sigma_\pi
\nuqe
as well as 
\equ
m_\pi(\ell_\pi;s_\pi,\ell_\pi) = 0
\uqe
The components $\ell_1$, $\sigma$ are parameterised by the single variable $s$  which we suitably scale to have the dimensions of length (and so allowing contact to be made with the PB theory). By construction $s$ will be zero if {\it both} $\sigma$ and $\ell_1$ vanish. A general linear parameterisation may be achieved using a free variable $\delta$ such that
\equn{l1D}
\ell_1 = s \sin \delta
\nuqe                                                     
and   
\equn{sigD}
\sigma = \kappa m_\alpha s \cos \delta
\nuqe
Dividing these we obtain 
\equ
\tan \delta = \frac{\ell_1 \kappa m_\alpha}{\sigma}
\uqe 
which establishes that the angle $\delta$ controls the relative magnitudes of the spin- and interfacial-like components. In particular if we set $\delta=0$ the collective coordinate is spin-like since there is no $\ell_1$ component and the constraint (\ref{GCon}) simply enhances the local magnetization. For this choice the field theory reproduces the $(\sigma,\ell)$ analysis discussed earlier. On the other hand $\delta = \frac{\pi}{2}$ corresponds to a purely interfacial collective coordinate in which $s$ denotes the position of the surface of fixed magnetization $m^{FJ}(z_1;\ell)$ and is essentially the same as the PB theory. Generally the collective coordinate has both spin-like and interfacial-like components satisfying
\equ
s^2 = \frac{\sigma^2}{\kappa^2 m_\alpha^2} + \ell_1^2
\uqe
which illustrates the metrical properties of the linear parameterisation (\ref{l1D}, \ref{sigD}). We will refer to such generalized fields as {\it proper} collective coordinates. The geometrical meaning of the proper collective coordinate and coupling angle is shown in Fig.\ \ref{euro}.
We note that it might be possible to choose different metrical factors (which scale the magnetization and the distance) to ensure that (\ref{sigD}) is dimensionally correct but the choice used here appears the most natural.

We could now proceed to construct a family of coupled Hamiltonians $\{ H[s,\ell;\delta] \}$ characterised by stiffness coefficients $\Sigma_{\mu \nu}(s,\ell;\delta)$ and binding potential $W(s,\ell;\delta)$. In fact we are also allowed to consider models in which $\delta$ varies with $\ell$. In this case the set of Hamiltonians whose properties we wish to study is denoted
\equ
{\cal H} = \{ H[s,\ell;\delta(\ell)] \}
\uqe
with elements distinguished by different choices of continuous function $\delta(\ell)$.

\subsection{The optimal model}
\label{theop}
The equivalence of each of the Hamiltonians in ${\cal H}$ as models of the MF free energy, adsorption and correlation functions certainly does not extend to fluctuation effects in $d=3$. As noted earlier there is no true non-trivial fixed point Hamiltonian for wetting at the upper critical dimension and it is possible that the different models in $d=3$ predict slightly different behaviour for some non-universal quantities. For example $H[s,\ell;0]$ or equivalently $H[\sigma,\ell]$ does not show any wetting parameter renormalization at complete wetting. To see this note that the arguments of the exponential terms in the binding potential (\ref{wsig1}) only depend on the upper field and substitution into the RG equations (\ref{RGdif}) leads to the same prediction for the value of $\omega$ determining $\theta$ as the CW and FJ models. This contrasts with the $H[s,\ell;\frac{\pi}{2}]$ model, equivalent to the PB model, which yields the different expression (\ref{barom}). In two dimensions however we can expect that all the models in ${\cal H}$ belong to the same universality class and flow to a fixed point under the action of the RG.  

These observations generate a host of questions. For example, why should we prefer the PB prediction for $\bar{\omega}$ over the $\omega$ result equivalent to CW theory? It is not sufficient to chose the PB description of complete wetting simply because it appears to fit the simulation results better at high temperatures. More generally we must ask which of the models in ${\cal H}$ best approximates the non-universality of the underlying LGW model? Recall that our hypothesis concerning the physical origin of problems (P1) and (P2) is that the CW model does not account for the coupling of interfacial fluctuations to the relatively small order parameter fluctuations near the wall. With this our task reduces to finding the coupling angle $\delta$ which most accurately models small order parameter fluctuations in a Gaussian approximation (which is the basis of the RG flow equation (\ref{RGdif})). We will argue that there is a unique answer to this question.         

Let us concentrate exclusively on the order parameter fluctuations at the wall. To this end consider a system with no CW-like fluctuations which may be achieved by {\it fixing} the location of the planar field $\ell_\pi$. Now ask what is the correlation function near the wall?  Because the fluctuations are small a Gaussian approximation certainly suffices in which case we are led to the prediction (using any of the models $H[s,\ell;\delta]$)
\equ
G^{\rm fix}(z_1,z_1;{\bf q}) = \left. \frac{ \Bigl( \frac{\partial m_\pi}{\partial s}(z_1;s,\ell_\pi) \Bigr)^2}{\frac{\partial^2 W}{\partial s^2} (s,\ell_\pi;\delta) + q^2 \Sigma_{11} (s,\ell_\pi;\delta) } \right|_{s=\tilde{s}=0}
\uqe
which by the invariance of the CFRS is independent of $\delta$. We may view this as the correct MF expression for the correlation function in a thin-film magnet of width $\ell$ with fixed spins $m^X=0$ at the $z=\ell$ layer (so that $G^{\rm fix}(z_2,z_2;{\bf q})=0$). Now focus on the value $\delta^*$ which maximises the equilibrium curvature: 
\equn{var1}
\frac{\partial^2 W}{\partial s^2}(s,\ell_\pi;\delta) < \frac{\partial^2 W}{\partial s^2}(s,\ell_\pi;\delta^*)  \mbox{\hspace*{5mm} for $\delta \ne \delta^*$}
\nuqe                                                         
with $s=0$. Note that in general $\delta^*$ will depend on $\ell$ as well as $T,h_1, h$ etc. Due to the invariance of the $G^{\rm fix}(z_1,z_1;{\bf q})$ with respect to $\delta$ the same value must also maximise the stiffness matrix element: 
\equn{var2}
\Sigma_{11}(0,\ell_\pi;\delta) < \Sigma_{11}(0,\ell_\pi;\delta^*)
\nuqe

The same argument implies that the variation of the order parameter $m_\pi$ with collective coordinate $s$ is also largest for this value:
\equn{var3}
\left| \frac{\partial m_\pi}{\partial s} \Bigl( z_1;s,\ell(\delta) \Bigr) \right| < \left| \frac{\partial m_\pi}{\partial s} \Bigl( z_1;s,\ell(\delta^*) \Bigr) \right| \mbox{\hspace*{5mm} for $s=0$}
\nuqe
Thus with this choice of coupling parameter $\delta=\delta^*$;
\setcounter{c}{0}
\begin{list}{(\roman{c})}{\usecounter{c} }
\item The Gaussian approximation is most accurate since the curvature is maximal. Recall that the fundamental assumption of the RG analysis is that a Gaussian approximation adequately describes the small fluctuations in the order parameter near the wall. In the effective Hamiltonian theory $H[s,\ell;\delta]$ the extent of these fluctuations is controlled by the curvature $\frac{\partial^2 W}{\partial s^2}(s,\ell;\delta)$. Ensuring that this is maximal avoids problems associated with the PB theory near $T=T_W$ when $\tilde{m}'_1$ vanishes.

\item The gradient approximation (i.e.\ the neglect of rigidity-like terms) implicit in the use of a Gaussian Hamiltonian for the lower field is most accurate. Importantly by maximising $\Sigma_{11}$ we completely avoid problems encountered in the PB theory when the lower stiffness vanishes --- the value of $\Sigma_{11}$ in the optimal theory is always finite. We also note that by maximising the stiffness we can also hope to have maximised the cut-off $\Lambda_1$ for the lower field which will no longer show any singular behaviour (at least for $T<T_C$). Therefore we have succeeded in taking any physics associated with the decoupling of fluctuations near $T=T_W$ out of the behaviour of the cut-off which had to be assumed in the PB theory. As we shall see the physics of decoupling effects associated with crossover from complete to critical wetting has a much more elegant description in the optimal model. 

\item The trace over $s$ accesses the largest region of the magnetization phase space since the variation of $m_\pi$ with $s$ is maximal. The simple geometrical interpretation of the optimal angle $\delta^*$ can be seen from Fig.\ \ref{euro} where it defines the normal to the FJ profile (in appropriate scaled units) at the point $z=z_1$.
\end{list}

It is instructive to compare the optimal angle $\delta^*$ with the worst possible choice of coupling angle which minimises the curvature, stiffness and variation $\frac{\partial m_\pi}{\partial s}$. In the geometrical interpretation this corresponds to directing $s$ along the tangent vector to $m^{FJ}(z;\ell_\pi)$ at $z=z_1$. Thus to first-order variation of $s$ produces no change in the profile so the constraint (\ref{GCon}) does not generate new configurations. Moreover it is easy to establish that the minimum values of $\frac{\partial^2 W}{\partial s^2}(0,\ell_\pi;\delta)$ and $\Sigma_{11}(0,\ell_\pi;\delta)$ are both zero. Thus for this choice of $\delta$ the Gaussian approximation completely breaks down.

In this way we propose that the best Hamiltonian in the set ${\cal H}$ for describing coupling between order parameter fluctuations near the interface and wall is
\equ
H[s,\ell] \equiv H[s,\ell;\delta^*]
\uqe
where $\delta^*$ satisfies the novel (and equivalent) variation conditions (\ref{var1})--(\ref{var3}). Moreover it is straightforward to derive a simple expression for the optimal coupling angle $\delta^*$ in terms of the known FJ profile. To do this note that the explicit formula for the variation $\frac{\partial m_\pi}{\partial s}$ is
\equ
\left. \frac{\partial m_\pi}{\partial s} (z_1;s,\ell_\pi) \right|_{s=0} = \kappa m_\alpha \cos \delta - \frac{\partial m^{FJ}}{\partial z} (z_1;\ell_\pi) \sin \delta
\uqe
which follows from differentiation of the constraint equation (\ref{GCon}) remembering (\ref{l1D}, \ref{sigD}). It is trivial to maximise the r.h.s.\ and hence derive
\equn{delta*}
\tan \delta^* = - \frac{ \frac{\partial m^{FJ}}{\partial z} (z_1;\ell)}{\kappa m_\alpha}
\nuqe
which is indeed the equation for the direction of the normal in Fig.\ \ref{euro}.

\subsection{Derivation of the position dependence}
While the fundamental equations for the construction of the optimal model $H[s,\ell]$ may appear to be rather involved and complicated, the actual derivation of the Hamiltonian requires only minor modifications to previous calculations. First we find the general expression for the Hamiltonian $H[s,\ell;\delta]$   with $\delta$ arbitrary then specify the optimal coupling angle via (\ref{delta*}).       

The constraint equation for the planar constrained profile is
\equn{planycon}
m_\pi(s_\pi \sin \delta; s_\pi, \ell_\pi) = m^{FJ}(0;\ell_\pi) + \kappa m_\alpha s_\pi \cos \delta
\nuqe
where we have set $z_1=0$ for convenience. It is a simple matter to perform the calculations with finite $z_1$ but our final results are unchanged provided we chose $0 \lesssim \kappa z_1 \le 1$. This may be viewed as the necessary mathematical condition which ensures that we are including the effect of order parameter fluctuations that are indeed near the wall. The planar constrained profile satisfies the Euler-Lagrange equation   
\equn{elagain}
\frac{\partial^2 m_\pi}{\partial z^2} (z; \cdots) = \phi' \Bigl( m_\pi(z; \cdots) \Bigr)
\nuqe                                                 
which we need to solve. As shown by FJ \cite{fjin} the all important position dependence of the various quantities is captured by the double parabola approximation \cite{lip3}
\equ
\frac{\partial^2 m_\pi}{\partial z^2} (z; \cdots) = \kappa^2_\alpha (m-m_\alpha) \mbox{\hspace*{5mm} for $z<\ell$}
\uqe
with $\phi(m)$ remaining arbitrary for $m<0$. Also note that position independent terms such as the interfacial stiffness coefficient(s) are not actually evaluated in any particular approximation but are left undetermined to be later combined into the appropriate dimensionless wetting parameter(s). The values of these are estimated by other means independent of the construction of the effective Hamiltonian. This is of course essential to the whole philosophy of effective Hamiltonian methods.   

The solution to (\ref{elagain}) is fully specified by the following boundary conditions

\begin{itemize}
\item For $0<z<s_\pi \sin \delta$
\equ
\left. \frac{\partial m_\pi}{\partial z} \right|_{z=0} = c m_{\pi 1}-h_1 
\uqe
and (\ref{planycon}).

\item For $s_\pi \sin \delta< z< \ell_\pi$
\equ
m_\pi(\ell_\pi;s_\pi,\ell_\pi) = 0
\uqe
and (\ref{planycon}).

\item For $z>\ell_\pi$ 
\eqa
m_\pi(\ell_\pi;s_\pi,\ell_\pi) &=& 0 \\
\stackrel{\rm lim}{\scs z \goto \infty} m_\pi(z;s_\pi,\ell_\pi) &=& m_\beta 
\aqe    
\end{itemize}

From the explicit solution for $m_\pi(z;s_\pi,\ell_\pi)$ it is straightforward (but tedious) to construct
\eqan{HPS}
H[s,\ell] &=& \int d{\bf y} \biggl\{ \frac{1}{2} \Sigma_{11} (\nabla s)^2 + \Sigma_{12} \nabla s . \nabla \ell \nonu \\
& & + \frac{1}{2} \Sigma_{22} (\nabla \ell)^2 + W_2(s,\ell;\delta) \biggr\}
\naqe
where the binding potential may be decomposed
\equ
W_2(s,\ell;\delta) = \frac{1}{2} r (s-s_0)^2 + W(s,\ell;\delta)
\uqe
Writing
\equ
m_1^{FJ} = m_\alpha + \tau + \Delta m_1^{FJ}(\ell)
\uqe
with
\equn{dmfj}
\Delta m_1^{FJ}(\ell) = - \frac{2\kappa m_\alpha}{c+\kappa} {\rm e}^{-\kappa \ell} - \frac{2\kappa \tau}{c+\kappa} {\rm e}^{-2\kappa \ell} 
\nuqe
we find (working to second-order in $\Delta m_1^{FJ}$)
\eqa
r &=& \kappa^2 (c+\kappa)q^2 \nonu \\
& & - \kappa (c+\kappa) \sin \delta \Bigl( 2(c-\kappa)q+\kappa \tau \sin \delta \Bigr) \Delta m_1^{FJ}(\ell)  \\
s_0 &=& (\kappa q)^{-1} \Delta m_1^{FJ}(\ell)  
\aqe
and (for $h=0$)
\eqan{HPSw}
\lefteqn{W(s,\ell)=} \nonu \\
& & 2\kappa m_\alpha \Bigl( m_\alpha \kappa s \cos \delta + \tau +  \Delta m_1^{FJ}(\ell) \Bigr) {\rm e}^{-\kappa(\ell-s \sin \delta)} \nonu \\
& & + \kappa \Bigl( m_\alpha^2 + (m_\alpha \kappa s \cos \delta + \tau )^2 \Bigr) {\rm e}^{-2\kappa(\ell-s \sin \delta)}
\naqe
where the useful abbreviation
\equ
q = m_\alpha \cos \delta + \tau \sin \delta 
\uqe
has been introduced. As usual we impose $W(s,\ell;\delta)=\infty$ for $\ell<s \sin \delta$.   

It is worthwhile pausing at this stage to emphasise the important features of the binding potential. First note that if we set $s=0$ then we automatically recover the FJ potential which neglects coupling effects. Secondly the form of the two-field potential is similar to that found in the PB model. In particular the exponential terms in the interaction term (\ref{HPSw}) only depend on the combined variable $\ell-s \sin \delta$ similar to the $\ell_2-\ell_1$ dependence in $H[\ell_1,\ell_2]$. If we set $\delta=\frac{\pi}{2}$ note that this variable indeed reduces to the difference $\ell_2-\ell_1$ consistent with our earlier statements that in this limit the proper field is interfacial-like. Similarly if we set $\delta=0$ the arguments of the exponential terms are independent of $s$ just as in the $H[\sigma,\ell]$ model confirming that for this case the proper field is spin-like.  

As can be anticipated the position dependence of the stiffness coefficients is a little more involved. We will only quote the results to the appropriate order that we will require in our calculations. As in the PB model the stiffness of the lower surface is essentially position independent and we do not incur any problems if we simply approximate
\eqan{HPSs11}
\Sigma_{11}(s,\ell;\delta) &\approx& \Sigma_{11}(0,\infty;\delta) \nonu \\
&=& \frac{\kappa q^2}{2} 
\naqe
Note that if we set $\delta=\frac{\pi}{2}$ we find $\Sigma_{11} \approx \frac{\kappa \tau^2}{2}$ which is in fact identical with the PB result (which of course it should be). On the other hand for $\delta=0$ the proper field is spin-like and the stiffness does not show any dependence on the scaling variable $\tau$. It is interesting to note that for small $\tau$ (\ref{HPSs11}) is the exact result found by solving (\ref{elagain}) for arbitrary potential function $\phi(m)$ i.e.\ beyond the double parabola approximation. This is because in the limit of $\tau \goto 0$ the profile near the wall is flat and the double parabola assumption does not lead to any error in the calculation.   

At leading order the off-diagonal element is
\equn{HPSs12}
\Sigma_{12}(s,\ell;\delta) \approx \kappa^2 m_\alpha q (\ell-s \sin \delta) {\rm e}^{-\kappa(\ell-s \sin \delta)}
\nuqe
and dominates the position dependence of the stiffness matrix ${\bf \Sigma}$ due to the algebraic pre-factor linear in $\ell-s \sin \delta$. For $\delta=\frac{\pi}{2}$ (\ref{HPSs12}) reduces to the PB expression and satisfies the stiffness matrix free energy relation (\ref{smfe2}). For general $\delta$ the analogue of this relation is not so elegant due to non-universal pre-factors which detract from the simplicity of the PB result. 

The stiffness coefficient for the surface of fixed magnetization $m^X=0$ is
\eqan{HPSs22}
\Sigma_{22}(s,\ell;\delta) &=& \Sigma_{\alpha \beta} + 2 \kappa m_\alpha( m_\alpha \kappa s \cos \delta + \tau) {\rm e}^{-\kappa(\ell-s \sin \delta)} \nonu \\
& &  - 2 m^2_\alpha \kappa^2 \nu (\ell-s \sin \delta) {\rm e}^{-2\kappa(\ell-s \sin \delta)} 
\naqe
showing the FJ-like negative next-to-leading order exponential term. Indeed setting $s=0$ recovers their result (\ref{fjstiff}) precisely.      

The final expression required to complete our specification of the optimal model is that for the temperature, surface field and position dependence of the optimal angle $\delta^*$. This can be easily calculated from the known FJ profile and we find
\equn{delta*2}
\tan \delta^* \simeq \frac{\tau}{m_\alpha} + \frac{2c}{c+\kappa} {\rm e}^{-\kappa \ell} + \frac{2 c \tau}{(c+\kappa) m_\alpha} {\rm e}^{-2\kappa \ell} + \cdots
\nuqe
which needs to be substituted into (\ref{HPSw})--(\ref{HPSs22}).

\section{Fluctuation effects in the optimal model}

\subsection{Complete wetting}   
The essential observation here is that in the approach to a complete wetting transition (route (iii) in Fig.\ \ref{pd}) the position dependence of the coupling angle is unimportant and we may write
\equn{5del}
\delta^* = \tan^{-1} \frac{\tau}{m_\alpha}
\nuqe
Within the $H[s,\ell]$ model it is the size of this coupling angle that describes the qualitative and quantitative nature of the coupling between the two fluctuating fields. First consider complete wetting transitions occurring at relatively high temperatures corresponding to $\tau \gg m_\alpha$. From (\ref{5del}) we see
\equ
\delta^* \approx \frac{\pi}{2}
\uqe
so that the proper field is interfacial-like. Thus deep in the complete wetting regime the optimal model is essentially the same as the PB model vindicating their earlier study. Indeed provided we keep away from the crossover region described below the PB model is a suitable starting point for the discussion of fluctuation effects. We can therefore anticipate on the basis of the optimal model that the value of the wetting parameter determining the critical amplitude $\theta$ is renormalized in accordance with the earlier coupled theory. We shall turn to this shortly.    

On the other hand as we consider complete wetting transitions occurring closer and closer to the critical wetting temperature we encounter a crossover regime in which the size of the coupling angle diminishes rather quickly. Again from (\ref{5del}) we see 
\equn{delta3}
\delta \approx \frac{\tau}{m_\alpha}
\nuqe
in this region ($\tau \ll m_\alpha$). Consequently near to the critical wetting temperature the proper field is essentially spin-like and the optimal model is basically the same as the $H[\sigma,\ell]$ Hamiltonian considered in Sec.\ \ref{interfacialand}. Recall our remarks made earlier that in the absence of any interfacial or translation component (so that the argument of the exponential terms in the binding potential only depend on the $\ell$ field) there is no increment to the effective value of the wetting parameter $\bar{\omega}$. Consequently due to the rotation of the coupling angle as the temperature is reduced we can also anticipate that 
\equ
\stackrel{\rm lim}{\scs T \goto T_W^+} \Bigl\{ \bar{\omega} \Bigr\} = \omega
\uqe
confirming the suggestion of PBS on the basis of the PB model.   

These remarks clarify the qualitative nature of the decoupling of fluctuations associated with the crossover from complete to critical wetting. The mechanism for this in the optimal model is far more elegant than the rather heuristic arguments given in Sec.\ \ref{problemswith} using the PB theory. Importantly we have succeeded in deriving a model of coupling effects which does not suffer from any pathological behaviour of the lower cut-off as $\tau \goto 0$. To see this note that the stiffness coefficient of the proper field in the optimal model is simply
\equ
\Sigma_{11} = \Sigma_{11}^W {\rm sec}^2 \delta^*
\uqe
where
\equn{sigw}
\Sigma_{11}^W = \frac{\kappa m_\alpha^2}{2}
\nuqe 
is the stiffness coefficient at the wetting temperature. Note that $\Sigma_{11}^W$ is perfectly well behaved as $T \goto T_W$  i.e.\ $\tau \goto 0$. Also recall our remarks made earlier when we noted that (\ref{sigw}) is the exact analytical MF result valid for arbitrary $\phi(m)$ and is not limited to the double parabola approximation.  

The main benefit of constructing the optimal model is seen by focusing on quantitative predictions for the scale of the wetting parameter renormalization in the crossover region (where (\ref{delta3}) is valid). To calculate the critical exponents and amplitudes for complete wetting it is sufficient to ignore the position dependence of the stiffness matrix elements and write the optimal model
\eqan{optRG}
H[s,\ell] & =& \int d{\bf y} \biggl\{ \frac{1}{2} \Sigma_{11} (\nabla s)^2 + \frac{1}{2} \Sigma_{22} (\nabla \ell)^2 + \frac{1}{2} r s^2 \nonu \\
& & + W(\ell - s \sin \delta^*) \biggr\}
\naqe
where $W(\ell)$ is the standard binding potential for wetting. This can be approximated by
\equ
W(\ell) = \bar{h} \ell + 2\kappa m_\alpha \tau {\rm e}^{-\kappa \ell} \mbox{\hspace*{9mm} for $\ell>0$}
\uqe
familiar from earlier approaches. Recall also that the parameters $\Sigma_{11}$ and $r$ are related to the transverse correlation length $\xi_{w\alpha}$ at the wall-$\alpha$ interface by $\xi_{w\alpha}^2 = \frac{\Sigma_{11}}{r}$. As has been emphasised before one of the satisfying features of the two-field approach is that the final predictions for observable critical amplitudes do not depend on terms like $\Sigma_{11}$ and $r$ separately but only on scaled combinations which can be related to physical quantities for other systems (such as the Ising model).     

By writing $\eta = s \sin \delta^*$ we see that the rescaled optimal model $H[\eta,\ell]$ and the PB Hamiltonian have precisely the same form so that we do not have to perform any new RG calculations. Consequently we can simply substitute the appropriate optimal model parameters into expression (\ref{barom}) derived by Boulter and Parry \cite{bp}. In particular in the crossover regime when the coupling angle is small we are led to the prediction
\equn{HPSRG}
\bar{\omega} = \omega + \Omega \frac{ (\tau/m_\alpha )^2}{1+(\Lambda \xi_{w\alpha})^{-2}} + O \left( ( \tau/m_\alpha)^4 \right)
\nuqe
where
\equn{Omega}
\Omega = \frac{k_BT \kappa^2}{4\pi \Sigma_{11}^W}
\nuqe
is the wetting parameter for the proper field evaluated at the wetting transition temperature. An important point to note here is that the cut-off $\Lambda$ appearing in our central result (\ref{HPSRG}) is perfectly well behaved. As we shall show in the next section the numerical value of $\Sigma_{11}^W$ is very similar to $\Sigma_{\alpha \beta}$ (at any temperature) so that we can reasonably anticipate that the cut-offs for both fields ($s$ and $\ell$) are the same. Alternatively we may go further and argue that the cut-off for the proper field is the same as the LGW model. This may arise asymptotically close to the wetting temperature as the proper field is then a spin-like variable. In the terminology of Sec.\ \ref{interfacialand} the spin-like Hamiltonians $\{ H[\sigma;z] \}$ can in principle be exactly calculated by evaluation of the functional integral (\ref{hsdef}) since $\sigma$  is always a single valued variable. Thus, in principle the models $\{ H[\sigma;z^X] \}$ carry the same information as the full LGW Hamiltonian concerning correlation functions along the spin-plane at $z=z^X$. This contrasts with interfacial variables for which the corresponding Hamiltonian is only appropriate for long wavelength fluctuations because overhangs in the field are ignored.  Fortunately these considerations are of no concern away from $T_C$ and for the most part we can safely assume that the cut-off for the $s$ and $\ell$ fields is the same as that present in the CW and FJ models.

Thus the optimized theory yields a precise perturbative expression for the surface field or temperature dependence of the renormalized wetting parameter determining the observable critical amplitude $\theta$. It follows that the increment to  $\theta$ vanishes quadratically as $\tau \goto 0^+$, a result that cannot be derived from the PB model. The coefficient of the quadratic term in (\ref{HPSRG}) is determined by the new wetting parameter $\Omega$ the value of which we need to estimate. 

\subsection{Critical wetting}
\label{critwetref}
In many respects the optimal model yields results for critical wetting that are supportive of the CW and FJ theories. However the model also allows a more precise expression of the queries raised at the end of Sec.\ \ref{problemswith} concerning the influence of coupling on the behaviour of local response functions near the wall. First let us state that if we neglect the position dependence of the stiffness coefficients then a lengthy but straightforward application of the renormalization group equations described in Sec.\ \ref{renorma} reveals the same results for the leading order critical behaviour as the CW model. In particular the critical exponent $\nu_\parallel$ for the correlation length $\xi_\parallel$ is strongly non-universal and dependent on the usual (un-renormalized) wetting parameter $\omega$. We do not repeat this analysis here since it is not a particularly illuminating calculation and the results can anticipated by inspection of the binding potential (\ref{HPSw}) with $\delta=0$. Similarly we can anticipate that the inclusion of the position dependence of the stiffness coefficients will drive the transition weakly first-order in the same way as the central result of the FJ theory. Actually this calculation requires a slight extension of the RG theory of two-field models since the scheme developed by Boulter and Parry (within linear and non-linear treatments of the interaction term $W(\ell_2-\ell_1)$) \cite{bp,bp2} outlined earlier does not cater for the presence of all the position dependent stiffnesses. Nevertheless such a scheme has recently been derived within a linear RG approximation to the interaction term and yields a very similar flow equation for the effective binding potential to the FJ equations \cite{bp5}. In particular it is no surprise to learn that the flows of $W$ and the $\Delta \Sigma_{\mu\nu}$  mix under RG iteration so that the critical wetting transition is destabilised due to presence of the FJ-like negative next-to-leading order exponential term in the $\Sigma_{22}$ stiffness coefficient. However as is now known the value of the tricritical wetting parameter $\omega^*$ at which the transition returns to being second-order is rather sensitive to the proper inclusion of the hard-wall contribution to the binding potential \cite{jinf1,B}. Consequently a detailed study of fluctuation-induced first-order behaviour in the optimal model requires a fully non-linear RG analysis including the position dependence of the stiffness matrix. Such a scheme has not been developed yet although we are aware of work in progress \cite{B2}. Consequently we content ourselves with the qualitative but none-the-less important remarks that the optimal model yields predictions for the true asymptotic critical singularities that are similar to the FJ and CW models.  

However there is behaviour associated with coupling effects that is not included in either of these models independent of the predictions for true asymptotic criticality. This is most simply seen by calculating the Ginzburg criterion for the surface susceptibility $\chi_1$ and comparing the result with that of the CW model (\ref{bh}). The Ginzburg criterion tests the self-consistency of the MF approximation and yields an estimate for the size of the correlation length when crossover from MF to non-classical behaviour occurs. Now this non-classical behaviour need not need be specified and may correspond to non-universality or indeed crossover to fluctuation-induced first-order effects. However we note that the estimates given by FJ for the size of the correlation length when the film jumps to infinity, characteristic of a first-order transition, are truly enormous and will turn out to be much larger than the crossover length that we calculate below. Consequently we feel justified in ignoring the position dependence of the stiffness matrix elements in our calculation.   

The calculation is for the susceptibility $\chi_1$ along the critical isotherm (route (ii) in Fig.\ \ref{pd}) and follows the same method as that outlined for the CW model by Halpin-Healey and Br\'{e}zin \cite{hb}. As $\tau$ is small we need to use the full expression for the position dependence of the coupling angle $\delta^*$  given in (\ref{delta*2}). The Hamiltonian is
\equ
H[s,\ell] = \int d{\bf y} \left\{ \frac{1}{2} \Sigma_{11}^W (\nabla s)^2 + \frac{1}{2} \Sigma_{\alpha \beta} (\nabla \ell)^2 + V(s,\ell) \right\}
\uqe
where we have written
\equ
V(s,l) = W_2 \Bigl( s,\ell;\delta^*(\ell) \Bigr) 
\uqe
and changed notation to avoid a plethora of $W$'s. Now from (\ref{delta*2}) we note that
\equ
{\rm e}^{-\kappa (\ell-s \sin \delta^*)} \approx {\rm e}^{-\kappa(\ell - s \frac{\tau}{p})} \left( 1 + O(s {\rm e}^{-\kappa \ell}) \right)
\uqe
where $p^2=m_\alpha^2 +\tau^2$. In this way (and using (\ref{dmfj}) and (\ref{HPSw})) we specify the explicit position dependence of the optimal model binding potential by
\eqan{W1}
\lefteqn{V(s,\ell) =} \nonu \\
& &  \frac{1}{2}\kappa^2 (c+\kappa) p^2 s^2 + a(1-s \frac{\tau}{p}) {\rm e}^{-\kappa(\ell - s \frac{\tau}{p})} \nonu \\
& &  + b {\rm e}^{-2\kappa(\ell - s \frac{\tau}{p})} + \bar{h}(\ell - s \frac{\tau}{p}) - \bar{h} \frac{2 c m^3_\alpha}{(c+\kappa)p^3} s {\rm e}^{-\kappa \ell}
\naqe
with
\eqa
a &=& 2\kappa m_\alpha \tau \\
b &=& \kappa (\tau^2 + \nu m_\alpha^2)
\aqe
and (as usual) $\nu=\frac{c-\kappa}{c+\kappa}$. Terms of order $O(s {\rm e}^{-2\kappa \ell},s^2 {\rm e}^{-\kappa \ell})$ have been ignored.    

Only a Gaussian approximation is used to include fluctuations and the onset of non-classical behaviour is heralded by the breakdown of the self-consistency of this approach. Expanding (\ref{W1}) about its minimum value gives
\eqa
V(s,\ell) &\simeq& V(\tilde{s},\tilde{\ell}) + \frac{1}{2} (\ell-\tilde{\ell})^2 \left. \frac{\partial^2 V}{\partial \ell^2} \right|_{\tilde{\ell},\tilde{s}} \nonu \\
& & +\frac{1}{2}(s-\tilde{s})^2 \left. \frac{\partial^2 V}{\partial s^2} \right|_{\tilde{\ell},\tilde{s}} 
\aqe
with of course the MF fields given by
\equn{mfcond}
\left. \frac{\partial V}{\partial \ell} \right|_{\tilde{\ell}} = \left. \frac{\partial V}{\partial s} \right|_{\tilde{s}} = 0 
\nuqe
Next we write the Hamiltonian in Fourier representation in anticipation of the functional integrals 
\eqa
H[s,\ell] &\simeq& \int \frac{d^2 {\bf q}}{(2 \pi)^2} \biggl\{ \frac{1}{2} (\Sigma_{11}^W q^2 + V_{ss}) |\hat{s}({\bf q})|^2 \nonu \\
& &  + \frac{1}{2} (\Sigma_{\alpha \beta} q^2 + V_{\ell \ell}) |\hat{\ell}({\bf q})|^2 \biggr\}
\aqe 
where we have written $\frac{\partial^2 V}{\partial n^2}=V_{nn}$ with all derivatives evaluated at $s=\tilde{s}$ and $\ell=\tilde{\ell}$. The partition function is
\equ
{\cal Z} = \int {\cal D}\ell {\cal D}s \: \exp(- H[s,\ell])
\uqe
and is evaluated in MF and Gaussian approximation. The MF result is
\equ
{\cal Z} \simeq {\rm e}^{-A V(\tilde{s},\tilde{\ell})}
\uqe
while in Gaussian approximation
\eqa
{\cal Z} & = & {\cal N} \int{\cal D}s {\cal D}\ell \; \exp \Biggl[ -\int \frac{d^2 q}{(2 \pi)^2} \biggl\{ \frac{1}{2} (\Sigma_{11}^W q^2 + V_{ss}) |\hat{s}({\bf q})|^2 \nonu \\
& & + \frac{1}{2} (\Sigma_{\alpha \beta} q^2 + V_{\ell \ell}) |\hat{\ell}({\bf q})|^2  \biggr\} \Biggr] \nonu \\
& = & {\cal N} \exp \Biggl[ \frac{A}{2} \int \frac{d^2q}{(2 \pi)^2} \biggl\{ \ln \Bigl[ \frac{\pi}{A^2(\Sigma_{11}^W q^2 + V_{ss})} \Bigr] \nonu \\
& & + \ln \Bigl[ \frac{\pi}{A^2(\Sigma_{\alpha \beta} q^2 + V_{\ell \ell})} \Bigr] \biggr\} \Biggr]
\aqe
where ${\cal N}$ is a suitable normalization factor which plays no role in the following calculation. In going between the last two equations we have made use of standard functional integrals \cite{ng}. Therefore the effective potential in Gaussian theory is
\eqa
V^{\rm eff}(\tilde{s},\tilde{\ell}) &=& V(\tilde{s},\tilde{\ell}) + \frac{1}{2} \int \frac{d^2 q}{(2 \pi)^2} \biggl\{ \ln(\Sigma_{11}^W q^2 + V_{ss}) \nonu \\
& & + \ln (\Sigma_{\alpha \beta} q^2 + V_{\ell \ell}) \biggr\}
\aqe
while the MF susceptibility is defined as
\equ
\chi_1 =  -\frac{\partial^2 V}{\partial h_1 \partial h} (\tilde{s},\tilde{\ell})
\uqe
the analogous Gaussian result is
\eqan{compare}
\chi_1^{\rm eff} & = & \chi_1 - \frac{1}{2} \frac{\partial^2}{\partial h \partial h_1} \int \frac{d^2 q}{(2 \pi)^2} \biggl\{ \ln(\Sigma_{11}^W q^2 + V_{ss}) \nonu \\
& & + \ln(\Sigma_{\alpha \beta} q^2 + V_{\ell \ell}) \biggr\} \nonu \\
& = & \chi_1 - \frac{1}{2} \int \frac{d^2 q}{(2 \pi)^2} \Biggl\{ \frac{ \frac{\partial^2 V_{ss}}{\partial h_1 \partial h} }{\Sigma_{11}^W q^2 + V_{ss}} - \frac{ \frac{\partial V_{ss}}{\partial h}.\frac{\partial V_{ss}}{\partial h_1}}{(\Sigma_{11}^W q^2 + V_{ss})^2} \nonu \\
& & + \frac{ \frac{\partial^2 V_{\ell \ell}}{\partial h_1 \partial h} }{\Sigma_{\alpha \beta} q^2 + V_{\ell \ell}} - \frac{\frac{\partial V_{\ell \ell}}{\partial h}.\frac{\partial V_{\ell \ell}}{\partial h_1} }{(\Sigma_{\alpha \beta} q^2 + V_{\ell \ell})^2} \Biggr\} 
\naqe
Evaluation of the partial derivatives in (\ref{compare}) is very algebraically intense and for our calculations we used the computer package MATHEMATICA. Once these are determined however the integrals are straightforward and we find
\eqa
\frac{\chi_1^{\rm eff}}{\chi_1} &=& 1 - \omega \left\{ \frac{1}{2} \ln(1+\xi_\parallel^2 \Lambda^2)+\frac{1}{1+\xi_\parallel^2 \Lambda^2} -1 \right\} \nonu \\
& & + \frac{\kappa^2}{4 \pi \Sigma_{11}^W} \ln (1 + \xi_{w\alpha}^2 \Lambda^2) + O(\xi_\parallel^{-4}) 
\aqe
where we have used the definitions $\xi_{w \alpha}^{-2} = \frac{W_{ss}}{\Sigma_{11}^W}$ and $\xi_\parallel^{-2} = \frac{W_{\ell \ell}}{\Sigma_{\alpha \beta}}$. The higher order terms are tiny in the crossover region and can be safely ignored. Following Halpin-Healey and Br\'{e}zin we suppose that a suitable definition of crossover is when
\equ
\left| \frac{\chi_1^{\rm eff}-\chi_1}{\chi_1} \right| = 1
\uqe
from which follows our central result
\eqan{ginzburg}
&\omega \left\{ \frac{1}{2} \ln(1+\xi_\parallel^2 \Lambda^2)+\frac{1}{1+\xi_\parallel^2 \Lambda^2} -1 \right\} = & \nonu \\
& 1 + \Omega \ln (1 + \xi_{w\alpha}^2 \Lambda^2) & 
\naqe

In the above expressions note that $\Lambda$ corresponds to the momentum cut-off for the $\ell$ and $s$ fields and may be safely identified with the cut-off for the CW and FJ models away from the bulk critical region.     

Importantly the optimal model expression for the Ginzburg criterion shows the presence of the wetting parameter $\Omega$ associated with the fluctuations of the proper field. Note that in the corresponding CW expression (\ref{bh}) there is no such term (which is equivalent to setting $\Omega=0$ in (\ref{ginzburg})) with the consequence that the true critical regime predicted by the present optimal coupled model is much smaller than previously expected. As we shall show in the next section the value of the second wetting parameter is about unity for Ising-like systems with the result that the crossover length $\xi_\parallel$ is very large. For this case we may further approximate (\ref{ginzburg}) by the expression
\equn{ginzbu}
\Lambda \xi_\parallel \approx {\rm e}^{1+\omega^{-1}} \left( 1 + \Lambda^2 \xi_{w\alpha}^2 \right)^{\frac{\Omega}{\omega}} + \cdots
\nuqe

Equations (\ref{HPSRG}), (\ref{ginzburg}) and (\ref{ginzbu}) are the main results of this paper. In the next section we turn to a discussion of the numerical value of the new wetting parameter $\Omega$ which will allow us to make better contact with the anomalous simulation data at the heart of problems (P1) and (P2).

\subsection{Value of the second wetting parameter $\Omega$}    
\label{valueof}   
Our final task is to specify the temperature dependence of the second wetting parameter $\Omega$ in order to fully quantify our central predictions (\ref{HPSRG}) and (\ref{ginzburg}). As with the CW and FJ theories we should regard $\Omega$ (and $\omega$) as inputs into the model which must be calculated by reliable means beyond MF approximation (at least in the bulk critical region). In particular we can identify two immediate concerns 

\setcounter{c}{0}
\begin{list}{(\roman{c})}{\usecounter{c} }
\item Establish a link between $\Omega$ and observable thermodynamic and correlation function properties of the underlying microscopic model.
\item  Determine the critical behaviour of $\Omega$ as $T_W \goto T_C^-$.  
\end{list}

We should emphasise here that these tasks stand somewhat apart from the rest of our paper and will probably require further work.

\subsubsection{Connection with thermodynamics}
 A link between the proper field wetting parameter $\Omega$ and thermodynamics follows from the CFRS. To see this note that we can use a proper field $s$ to describe the surface correlations at the wall-$\alpha$ interface --- equivalent to setting $\ell=\infty$ in the Hamiltonian $H[s,\ell]$ (with $h=0$ of course). According to the CFRS 
\equ
G^{w\alpha}(0,0;{\bf q}) = \left. \frac{ \left( \frac{\partial m_\pi}{\partial s} (0;s,\infty) \right)^2 }{\frac{\partial^2 W}{\partial s^2}(s,\infty;\delta) + q^2 \Sigma_{11}(s,\infty;\delta) + \cdots} \right|_{s=0}
\uqe                                                  
and is valid for any value of $\delta$. Now exactly at the wetting transition temperature the optimal angle $\delta^*=0$ (with $\ell=\infty$) and using this we find
\equ
G^{w\alpha}(0,0;{\bf q}) = \frac{ m_1^2 \kappa^2}{r+q^2 \Sigma_{11}^W}
\uqe
using the fact that the surface and bulk magnetizations are equal at $T=T_W$ and $h=0$. From an exact sum-rule identity \cite{ep} we can relate the zeroth moment $G_0(0,0)$ to the wall susceptibility 
\equ
G_0(0,0) = \frac{\partial m_1}{\partial h_1} \equiv \chi_{11}/k_BT
\uqe
and also recall that $\Sigma_{11}^W$ and $r$ determine the (second-moment) correlation length $\xi_{w\alpha}$. Consequently we can write
\equ
\frac{\Sigma_{11}^W}{k_BT} = \frac{m_1^2 \kappa^2 \xi^2_{w \alpha}}{\chi_{11}^{(\alpha)}}
\uqe 
and bare in mind that quantities are defined for the wall-$\alpha$ interface. In this way the value of the new wetting parameter (at a particular wetting temperature) can be identified as
\equn{Omega2}
\Omega = \frac{\chi_{11}^{(\alpha)}}{4\pi m_1^2 \xi_{w \alpha}^2} \mbox{\hspace*{10mm} for $T=T_W$}
\nuqe
In principle all the quantities on the r.h.s.\ can be precisely determined for the full LGW model.   

The argument given above is not completely rigorous since it is based on reinterpretation of the CFRS equations which has only been established at MF level. Thus one may legitimately worry therefore whether (\ref{Omega2}) is only appropriate away from the bulk critical region. In case this query is well founded we will content ourselves that the final identification above is reliable outside the bulk critical regime and try a different method of attack for $T_W \lesssim T_C$.

\subsubsection{Universality of $\Omega$}
Here we give an argument aimed at establishing the universality of the parameter $\Omega$ as $T \goto T_C^-$ i.e.\ for wetting transitions occurring close to the bulk critical point. In fact the argument leads to a rather precise estimate for the critical value
\equ
\Omega_C = \stackrel{\lim}{\scs T \goto T_C} \Bigl\{ \Omega(T,\cdots) \Bigr\}
\uqe
complementing the known universality of $\omega(T_C)$. The reasoning we give is different to the simpler one used earlier \cite{parryrev} although the final numerical estimates are rather similar.   

In the bulk critical region the surface correlation length $\xi_{w\alpha}$ is very large and the simpler expression (\ref{ginzbu}) for the Ginzburg criterion is appropriate. However in this result the two wetting parameters do not appear separately but as a ratio and it is natural to focus on this combination. We noted before that expression (\ref{sigw}) for $\Sigma_{11}^W$ was exact in MF theory for arbitrary $\phi(m)$. We can extend this by including an arbitrary pre-factor $f_2(m)$ (denoted $K$ in \cite{sp}) as the coefficient of $\frac{1}{2} (\nabla m)^2$ in the LGW Hamiltonian. With this the exact result for the lower stiffness of the proper field at $T_W$ becomes
\equn{sigw2}
\Sigma_{11} = \frac{f_2(m_\alpha) m^2_\alpha \kappa}{2}
\nuqe
Now following the pioneering studies of Fisk and Widom \cite{fw} (reviewed in some detail by Rowlinson and Widom \cite{rowlinson}) we can chose $f_2$ and $\phi(m)$ to mimic the inclusion of bulk critical fluctuations in a variational (i.e.\ MF) treatment of the functional $H_{LGW}[m]$. This appears, of course, in the saddle point identification used in constructing effective Hamiltonians. From the precise variational result (\ref{sigw2}) and the definition of $\omega$ it is straightforward to show that the critical ratio
\equ
\frac{\Omega}{\omega} = 8 K \mbox{\hspace*{10mm} $T_W \rightarrow T_C$}
\uqe
where $K$ is the surface tension critical amplitude defined in (9.82) of \cite{rowlinson}.   

The quantity $K$ naturally arises in the Fisk-Widom theory of the surface tension and its value has been estimated using a variety of techniques. Rowlinson and Widom quote the predictions $K=\frac{1}{6}$ and $K\approx \frac{1}{7}$ arising from pure MF and the rather accurate Fisk-Widom theory respectively. From a fundamental point of view $K$ is a nice quantity because it can be calculated using the celebrated $\epsilon$ expansion in the RG analysis of the LGW model. Therefore directly from the known first-order result in $\epsilon$ we find \cite{ok}
\equn{eexp}
\Omega_C = \frac{4}{3} \omega_C \left[ 1-\left( \frac{\sqrt{3} \pi}{9} - \frac{1}{2} \right) \epsilon + O(\epsilon^2) + \cdots \right]
\nuqe
evaluated at $\epsilon=1$. Thus we predict
\equn{Omegaresult}
\Omega_C \approx 0.92
\nuqe
close to the Fisk-Widom result $\Omega_C \approx \frac{8}{7} \omega_C$. The error in this estimate is small (and in the last digit) because of the accuracy of $\omega_C$ and $K$.  

The argument given above is different to one used earlier \cite{parryrev} which was based on a direct analysis of (\ref{sigw2}) using Fisk-Widom theory to express $f_2$ in terms of bulk thermodynamic functions. This also predicts that $\Omega_C$ is universal but the final expression is solely in terms of bulk critical amplitudes and looks rather different to (\ref{eexp}). Interestingly however the final numerical estimate is rather close to (\ref{Omegaresult}) although we prefer the present argument because it reduces $\Omega_C$ to previously studied surface quantities.

\subsection{Connection with simulations}
\subsubsection{Critical wetting; (P1) revisited}
Let us denote the solutions to (\ref{bh}) and (\ref{ginzburg}) determining the respective size of the MF regime in CW and coupled models respectively by $\xi_{\rm Gi}^{CW}$ and $\xi_{\rm Gi}$. Then we see that the effect of coupling is to dramatically {\it reduce} the extent of the true asymptotic regime for the susceptibility $\chi_1$  since
\equ
\frac{\xi_{\rm Gi}}{\xi_{\rm Gi}^{CW}} \approx \left( 1+\Lambda^2 \xi_{w\alpha}^2 \right)^{\frac{\Omega}{\omega}}
\uqe
It is easiest to assume that the exponent on the r.h.s.\ is close to the critical value $\frac{\Omega_C}{\omega_C} \approx \frac{8}{7}$. Alternatively we may argue that away from $T_C$ the exponent is determined by the MF value of $K$ in which case we can estimate $\frac{\Omega}{\omega} \approx \frac{4}{3}$. We do not know the precise value of the surface correlation length $\xi_{w\alpha}$ in the simulation studies but a sensible estimate away from the bulk critical regime is $\Lambda \xi_{w\alpha} \approx \pi$ leading to
\equ
\frac{\xi_{\rm Gi}}{\xi_{\rm Gi}^{CW}} \approx 15 - 24
\uqe                                                    
depending on the choice of $K$. The coupled model leads to the same prediction for the divergence of $\xi_\parallel$ as the CW model and we can use the estimate of Halpin-Healey and Br\'{e}zin \cite{hb} 
\equ
\xi_\parallel \Lambda \approx \frac{0.15}{\sqrt{\frac{h}{J}}}
\uqe
for the Ising model with $T_W \approx 0.63 T_C$. Here $J$ is the usual spin-spin coupling strength. In this way we predict crossover occurs when
\equ
\frac{h}{J} \sim 10^{-6}
\uqe
(compared to the CW result $\sim 10^{-4}$) which is much smaller than the values studied in the simulations.    

Thus the optimised theory of coupling effects yields a quantitative explanation of (P1) and allows a precise expression of the conjectures made earlier on the basis of the PB model. As we have emphasised before future simulation studies should focus on the measurement of response functions local to the $\alpha \beta$ interface if they hope to see fluctuation effects consistent with the original CW model predictions for critical wetting.

\subsubsection{Complete wetting; (P2) revisited}
As regards the BLF simulations \cite{blf,blf1} the most important results of the optimal coupled model are the qualitative remarks

\setcounter{c}{0}
\begin{list}{(\roman{c})}{\usecounter{c} }
\item For $T \gg T_W$ the PB model is recovered so that the effective value of the wetting parameter $\bar{\omega}$ is greater than $\omega$. Hence the value of $\theta> 1+\frac{\omega}{2}$.

\item As $T \goto T_W^+$ the increment to $\bar{\omega}$ vanishes so that the extrapolated value of $\bar{\omega}$  (or $\theta$) at $T_W$ is the CW result. Thus the optimised model is entirely supportive of the prediction (\ref{wlimit}) of PBS \cite{pbs} based on the PB model.
\end{list}

It is worth re-emphasising the merits of extracting the value of $\omega(T_W)$ by extrapolating from the data for $\theta$ not least because it yields a value for $\omega \approx 0.8$ consistent with long standing theoretical predictions. Importantly it avoids the issue of whether the wetting transition is weakly first-order or not. Moreover the BLF data are taken from susceptibility measurements near to the $\alpha \beta$ interface (and not at the wall) where fluctuation effects are very strong. The fact that the BLF simulation results are in quantitative agreement with these predictions is (currently) the most important independent confirmation of the hypothesis concerning coupling effects.   

Unfortunately the BLF study is not sufficiently accurate to test the quantitative prediction (\ref{HPSRG}) although the data is certainly consistent with a quadratic power law for $\bar{\omega}$ near $T_W$ (see Fig.\ \ref{bigtheta}). The small $\delta^*$ result simplifies further if we consider complete wetting transitions occurring close to the bulk critical point. If $T$ is fixed close to $T_C$ and $h_1 \gtrsim h_1^W$, then provided that $\Lambda \xi_{w\alpha}$ and $c/ \kappa$ are large enough we can write
\equ
\bar{\omega} = \omega_c + \Omega_C \left( \frac{h_1-h_1^W}{h_1^W} \right)^2 + \cdots
\uqe
which is a universal result. We accept here that forcing our theory to work in the vicinity of the bulk critical point is a contentious issue since there are well known difficulties interpreting effective Hamiltonian models in the bulk critical region \cite{huse}. Therefore we wish to not dwell too long on the nature of $\bar{\omega}$ near $T_C$.

\subsubsection{A conjecture for the value of $\omega_C$}   
Before we close the paper we comment on an intriguing numerical coincidence that caught our attention when estimating the extent of the critical region for $\chi_1$ on the basis of the coupled theory. If we focus on the Ginzburg result (\ref{ginzbu}) in the bulk critical region we are naturally led to the evaluation of the universal number
\equ
{\rm e}^{1+\omega_C^{-1}}
\uqe
Taking the Fisher-Wen estimate \cite{fishw} $\omega_C \approx 0.77_5$ we find that this quantity is extremely close to $\pi^2$! It is therefore tempting to conjecture that the exact value of $\omega_C$ is
\eqa
\omega_C &=& \frac{1}{2 \ln \pi -1} \nonu \\
&=& 0.7755186
\aqe
although this is of course entirely speculative.

\section{Discussion}  
We begin with a summary before addressing some pertinent questions.

\subsection{Summary of method and results}
\label{summaryof}
The starting point of our study is the coupling hypothesis
\begin{quote}
Some of the non-universal physics of wetting transitions at the upper critical dimension (for systems with short-range forces) $d=3$ is sensitive to the coupling between order parameter fluctuations near the $\alpha \beta$ interface and wall.
\end{quote}
The rest of our analysis is a thorough investigation of the procedures we must follow in order to construct a two-field model of coupling effects at critical and complete wetting transitions. We first address (P3) related to the reconstruction of MF correlations and establish the following result:      
\begin{quote}
For arbitrary choices of local collective coordinates $X_\mu({\bf y})$, the two-field model $H[X_1,X_2]$ (defined using a FJ-like saddle point identification) can precisely recover the order parameter correlations of the underlying LGW theory (in MF approximation) using the invariant relation
\eqa
\lefteqn{G(z_\mu,z_\nu;{\bf q}) =} \nonu \\
& &  \frac{\partial m_\pi}{\partial X_{\mu'}} (z_\mu;\{ X_i \}) \frac{\partial m_\pi}{\partial X_{\nu'}} (z_\nu; \{ X_i \} ) S_{\mu' \nu'}({\bf q}) 
\aqe
where $z_1$, $z_2$ are the MF locations of the two fields $X_1, X_2$ and ${\bf S}$ is the structure factor matrix.
\end{quote}

To consider fluctuation effects beyond MF we must carefully chose the collective coordinates $X_\mu({\bf y})$. Following all earlier studies we adopt a standard interfacial variable $X_2=\ell$ for the upper field which we define as the surface of fixed magnetization $m^X=0$. The choice of $X_1$ is not so obvious and we therefore focus on a set of Hamiltonians
\equ
{\cal H} = \{ H[s,\ell;\delta] \}
\uqe
characterised by different coupling angles $\delta$ which describe the relative importance of the spin-like ($\sigma$) and interfacial-like ($\ell_1$) components of a proper field $s$ with magnitude satisfying
\equ
s^2 = \ell_1^2 + \frac{\sigma^2}{m_\alpha^2 \kappa^2}
\uqe

At the upper critical dimension the Hamiltonians in the set ${\cal H}$ can exhibit different non-universal critical behaviour (for some quantities). We argue that the model,
\equ
H[s,\ell] = H[s,\ell;\delta^*]
\uqe
with $\delta^*$ the optimal coupling angle, best describes the fluctuations of the order parameter near the wall as determined by a novel variational principle. Analysis of the optimal model yields the following predictions

\setcounter{c}{0}
\begin{list}{(\arabic{c})}{\usecounter{c} }
\item Deep in the complete wetting regime the proper field is interfacial-like and the model reduces to the PB theory. RG analysis predicts the same wetting parameter renormalization effect at high temperatures and an elegant crossover mechanism associated with the rotation of $\delta^*$ as the temperature is reduced to $T_W$. This leads to the explicit perturbation expansion result
\equn{HPSRG2}
\bar{\omega} = \omega + \Omega \frac{ (\tau/m_\alpha )^2}{1+(\Lambda \xi_{w\alpha})^{-2}} + \cdots
\nuqe
in the vicinity of $T_W$.

\item At critical wetting the effect of coupling is to dramatically reduce the size of the true asymptotic critical regime for the local susceptibility $\chi_1$ at the wall. Specifically crossover to non-classical behaviour should occur when the correlation length is
\equ
\Lambda \xi_\parallel \approx {\rm e}^{1+\omega^{-1}} \left( 1 + \Lambda^2 \xi_{w\alpha}^2 \right)^{\frac{\Omega}{\omega}}
\uqe
much larger than corresponding CW result.

\item The new physics that emerges is controlled by a second wetting parameter $\Omega$ that has no counterpart in simpler one-field theories. We argue that $\Omega$ approaches a universal value $\Omega_C \approx 0.92$ in the bulk critical regime.

\item The values of the critical exponents in the true asymptotic critical regimes of critical and complete wetting are the same as those of the CW theory. Similarly it is highly likely that the optimal model exhibits fluctuation-induced first-order behaviour for sufficiently small $\omega<\omega^*$ similar to the FJ theory.
\end{list}

\subsection{Remaining questions}
While one can argue that the coupling hypothesis is justified {\it a posteriori} it is still perhaps surprising that the inclusion of coupling effects leads to the modification of the CW model results in (\ref{HPSRG2}). To finish therefore we try to stand back from the mathematics and give more physical answers to some important questions. We begin with

\subsubsection{Why do coupling effects appear to matter so much?}
Firstly we should emphasise that most of the physics is insensitive to coupling effects. As mentioned in the last remark of Sec.\ \ref{summaryof} even though we are at the upper critical dimension the values of the critical exponents for both critical and complete wetting are identical to those of the CW model. Nevertheless one can force the question and ask why does any of the underlying physics depend on the coupling to a weakly fluctuating field? To answer we draw on recent work aimed at studying wetting at rough walls \cite{psf}.

If we fix the position of the lower field $\ell_1({\bf y}) = z_W({\bf y})$ (say) in the PB model the two-field Hamiltonian $H[z_W,\ell_2]$ describes the problem of wetting at a fixed corrugated non-planar wall. However even at MF level it is now known that the surface phase diagram is effected strongly by minor deviations of the wall from the plane. In particular analytical and numerical studies of a LGW model show that a (planar) critical wetting transition is driven first-order (and occurs at a lower temperature) provided the width $a$ of the walls deviation satisfies \cite{psf}
\equ
a > \sqrt{\frac{c-\kappa}{c+\kappa}} \kappa^{-1}
\uqe
which is less than a bulk correlation length. More generally numerical studies \cite{stella} of CW-like models of wetting at rough walls in $d=2$ shows that the order of the wetting transition is altered if
\equ
\zeta_S \ge \zeta
\uqe
where $\zeta$ is the roughness exponent for the free $\alpha \beta$ interface and $\zeta_S$ is the roughness exponent for the wall configurations. Now let us return to the case of wetting at planar walls but include the coupling of order parameter fluctuations using a two-field model. It is natural to make analogy between the non-planar wall and fluctuations near a planar wall which increase the effective area as seen by the upper surface. Since the fluctuations in the order parameter near the wall are small they can only correspond to some effective roughness $\zeta_S=0$. Thus we surmise that it is possible that fluctuation effects are sensitive to the coupling if
\equ
0 \ge \frac{3-d}{2}
\uqe
which is only marginally satisfied in $d=3$. Recall that for systems with long-range forces MF theory is expected to be valid in this dimension which suggests that coupling can only be of any possible importance for those with short-range interactions. We also remark that in $d=2$ transfer matrix studies confirm that coupling does not change any of the universal critical properties associated with wetting transitions. We do not include the details of the calculation here \cite{pa2} since we feel it would take us away from our central concerns.

These remarks complement those made earlier \cite{bp2} that in the non-linear RG analysis of the CW model \cite{lf} a shift in the origin of the binding potential (i.e.\ the location of the wall) is a marginal operation in $d=3$ (and irrelevant for $d<3$). Since such fluctuations are allowed for in two-field theory (by the variation of $s \sin \delta^*$) it is perhaps not surprising that some of the non-universal physics is sensitive to coupling effects.

Hopefully the above discussion helps clarify the reason why fluctuation effects can be sensitive to coupling at the upper critical dimension. If we accept their importance it is natural to proceed to our final question

\subsubsection{How should we visual coupled fluctuations?}

In order to address this question we first recall the answer to a related one --- what fluctuations give rise to the standard binding potential $W(\ell)$ in the CW model? In the absence of long-range wall-fluid and fluid-fluid forces the direct interaction between the interface and wall also arises due to fluctuation effects. This issue has been emphasised by a number of authors notably Gompper {\it et.\ al.\ } \cite{gkl} who point out that within a solid-on-solid (SOS) approximation an effective binding potential arises due to the presence of spike-like fluctuations from the interface to the wall (which can be seen in simulation studies). 
\begin{figure}[h]
\begin{center}
\scalebox{0.6}{\includegraphics{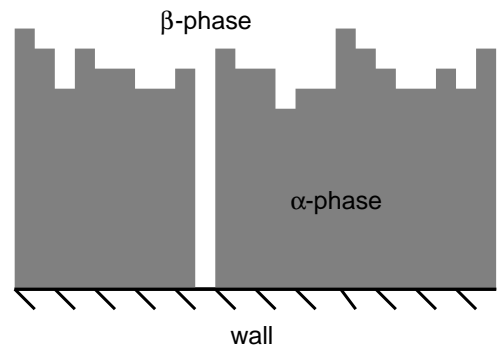}}
\end{center}
\caption{Example of a spike-like fluctuation in the SOS model carrying information from the wall to the fluid interface. Such configurations are believed to give rise to the exponentially decaying terms in the binding potential.}
\label{bendsos}
\end{figure}
Note that these are not particularly important in $d=2$ because of the contact condition $\ell \sim \xi_\perp$. 

\begin{figure}[h]
\begin{center}
\scalebox{0.55}{\includegraphics{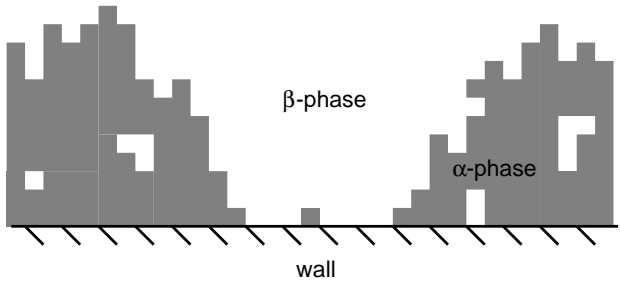}}
\end{center}
\caption{In $d=2$ tube or spike-like fluctuations are no longer needed to carry information between the wall and the fluid interface as the contact condition $\ell \sim \xi_\perp$ ensures that the interface itself regularly visits there.}
\label{bend2}
\end{figure}

\begin{figure}[h]
\begin{center}
\scalebox{0.6}{\includegraphics{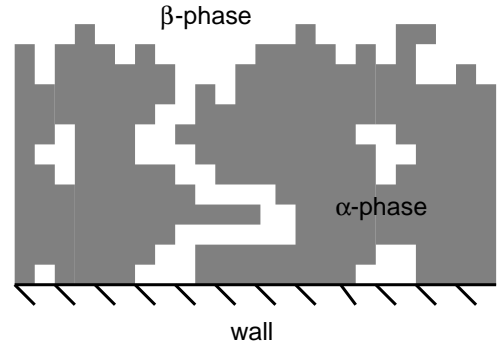}}
\end{center}
\caption{Tube-like fluctuation from the interface to the wall in an Ising-like system. The tube is deformed due to its interaction with a local droplet-like fluctuation near the wall typical of fluctuations at the wall-$\alpha$ interface. Those droplet-like fluctuations away from the wall or near the $\alpha \beta$ interface only result in a renormalization of the MF expressions for $\kappa$ and $\Sigma_{\alpha \beta}$ respectively.}
\label{bend3}
\end{figure}

In $d=3$ however where  $\ell \gg \xi_\perp$, the interface rarely reaches the wall and the spike-like fluctuations are crucial to the picture, determining (at least in part) the structure of $W(\ell)$. Beyond the SOS approximation we can expect the spikes to correspond to string or tube-like objects of $\beta$-like magnetization that wander on their way from the $\alpha \beta$ interface to the wall. This is reminiscent of the Abraham-Chayes-Chayes (ACC) description of bulk correlations in the $d=3$  Ising model which is known to reproduce the expected Ornstein-Zernike decay of the correlation function \cite{acc}. In this interpretation the leading order exponential decay of the binding potential can be directly related to bulk-like correlations (i.e.\ wandering SOS tubes) between points near the $\alpha \beta$ interface and wall \cite{abraham}. The coupled model describes deformations to the ACC tube arising from interactions with local (droplet-like) fluctuations  which increase the effective area of the wall. This is consistent with the heuristic scaling theory discussed in Sec.\ \ref{interp} that coupling alters the manifestation of the {\it direct} interaction between the fluctuating surfaces. These ideas are illustrated schematically in Figs.\ \ref{bendsos}, \ref{bend2}, \ref{bend3} and only constitute a minor change to the overall picture implicit in the CW model. 
Moreover with this interpretation it is clear why in $d=2$ coupling is not important because the spike-like fluctuations are not needed to convey information from the wall to the interface. Nevertheless we surmise that coupled models are still applicable even if they do not change any of the critical properties associated with wetting transitions as such theories should provide a more reliable description of correlation function structure.

With this picture it is also clear why it is not necessary to explicitly include more fields describing fluctuations at other positions since such effects are implicitly included in the effective Hamiltonian theory when we replace the MF $\kappa$ and $\Sigma_{\alpha \beta}$ by their {\it true} Ising values (see Fig.\ \ref{bend3}).

\acknowledgements
The authors are indebted to Dr C.\ J.\ Boulter for discussions on much of the work described here. This research was supported by the Engineering and Physical Science Research Council of the United Kingdom.

\end{document}